\documentclass[11pt,a4paper]{article}

\usepackage[T1]{fontenc}

\usepackage{jheppub}
\usepackage{graphicx}
\usepackage{amsmath}
\usepackage{xspace}
\usepackage[dvipsnames]{xcolor}
\usepackage{float}
\usepackage{comment}
\usepackage{subcaption}
\usepackage{textcomp}
\usepackage{placeins}
\usepackage{siunitx}
\usepackage{hhline}
\usepackage[normalem]{ulem}

\newcommand{\eq}[1]{eq.~\eqref{eq:#1}}
\newcommand{\eqs}[2]{eqs.~\eqref{eq:#1} and \eqref{eq:#2}}
\renewcommand{\sec}[1]{sec.~\ref{sec:#1}}

\newcommand{\subsec}[1]{sec.~\ref{subsec:#1}}
\newcommand{\fig}[1]{fig.~\ref{fig:#1}}

\newcommand{\figs}[2]{figs.~\ref{fig:#1} and \ref{fig:#2}}
\newcommand{\app}[1]{App.~\ref{app:#1}}
\newcommand{\apps}[2]{Apps.~\ref{app:#1} and \ref{app:#2}}
\newcommand{\tab}[1]{Tab.~\ref{tab:#1}}

\newcommand{\abs}[1]{\lvert#1\rvert}
\newcommand{\ord}[1]{\mathcal{O}(#1)}

\newcommand{\df}{\mathrm{d}}

\newcommand{\as}{\alpha_{\rm s}}

\newcommand{\Ecm}{E_{\mathrm{cm}}}

\newcommand{\Tau}{\mathcal{T}}
\newcommand{\Taufr}{\mathcal{T}_2^{\mathrm{FR}}}

\newcommand{\GeV}{\,\mathrm{GeV}}

\newcommand{\nn}{\nonumber}

\newcommand{\cP}{\mathcal{P}}

\newcommand{\cut}{\mathrm{cut}}

\newcommand{\FO}{\mathrm{FO}}

\newcommand{\NLO}{\mathrm{NLO}}

\newcommand{\NNLL}{\mathrm{NNLL}}

\newcommand{\nons}{\mathrm{nons}}

\newcommand{\one}{{(1)}}

\newcommand{\twoj}{\mbox{$2$-jettiness}\xspace}
\newcommand{\threej}{\mbox{$3$-jettiness}\xspace}
\newcommand{\nj}{\mbox{$N$-jettiness}\xspace}

\newcommand{\lqcd}{\Lambda_\mathrm{QCD}}

\newcommand{\dgamMC}{\df\Gamma^\textsc{mc}}

\newcommand{\geneva}{\textsc{Geneva}\xspace}

\newcommand{\amcatnlo}{\textsc{aMC@NLO}\xspace}

\newcommand{\minlo}{\textsc{MiNLO}\xspace}

\newcommand{\pythia}{\textsc{Pythia}\xspace}
\newcommand{\pythiaEight}{\textsc{Pythia8}\xspace}

\renewcommand{\thesection}{\arabic{section}}

\allowdisplaybreaks[2]

\newcommand{\rescalethreeplots}{0.32\textwidth}
\newcommand{\hspacebetweenthreeplots}{-1ex}

\newcommand{\spacebeforefigurecaption}{-1ex}

\begin{document}


\preprint{ZU-TH 35/20}

\title{Resummed predictions for hadronic Higgs boson decays}
\author[a]{Simone Alioli,}

\author[a]{Alessandro Broggio,}

\author[a]{Alessandro Gavardi,}

\author[a \quad]{Stefan Kallweit,}

\author[a]{Matthew A.~Lim,}

\author[a]{Riccardo Nagar,}

\author[a]{Davide Napoletano}

\author[a,b]{and Luca Rottoli}

\affiliation[a]{Universit\`{a} degli Studi di Milano-Bicocca \& INFN, Piazza della Scienza 3, Milano 20126, Italia\vspace{0.5ex}}

\affiliation[b]{Physik Institut, Universit\"at Z\"urich,
  Winterthurerstrasse 190, 8057 Z\"urich, Switzerland}

\emailAdd{simone.alioli@unimib.it, alessandro.broggio@unimib.it, a.gavardi@campus.unimib.it, stefan.kallweit@unimib.it, matthew.lim@unimib.it, riccardo.nagar@unimib.it, davide.napoletano@unimib.it, luca.rottoli@physik.uzh.ch}

\date{\today}

\abstract{

We present the NNLL$'$ resummed \twoj distribution for decays of the
Standard Model Higgs boson to a $b\bar{b}$-quark pair and to
gluons. The calculation exploits a factorisation formula derived using
Soft-Collinear Effective Theory, in which large logarithms of the
\twoj are resummed by renormalisation group evolution of the 
hard, soft and jet contributions to the differential decay rate. We
match the resummed predictions to the fixed-order NNLO result using
the \geneva framework, extending the validity of the results to all
values of the resolution variable and providing a fully exclusive NNLO
event generator matched to the \pythiaEight parton shower.  }

\maketitle
\flushbottom

\section{Introduction}
\label{sec:intro}
The lack of any definitive signal of New Physics at the Large Hadron
Collider (LHC) suggests that the high-energy physics community must be open to alternative ways to
probe Beyond the Standard Model (BSM) effects at collider
experiments. 
In particular, precision measurements of the Standard Model (SM) Higgs sector
may be a way to indirectly constrain BSM theories which live at scales
beyond our reach, both at the LHC and at future lepton colliders or
`Higgs factories'. It is therefore crucial that theoretical
predictions for processes involving the production or decay of the
Higgs boson have a precision which matches that of experiment.

In this work we consider the hadronic decays of a Higgs
boson to a $b\bar{b}$-quark pair and to gluons.\footnote{
 Neglecting light quark Yukawa couplings, these decay channels
 constitute the only hadronic channels in the Standard Model at this perturbative order~\cite{Davies:2017xsp}. Indeed, even if one were to consider decays to lighter quarks the extension from the $b$-quark case implemented here would be relatively trivial since we treat them as kinematically massless.} Although these modes
are difficult to observe at a hadron
collider due to the large QCD background, they may well prove to be
useful probes of Higgs physics at a future lepton collider since they
contribute significantly to the total width of the Higgs
boson.

The dominant decay mode of the Higgs boson is to a $b\bar{b}$-quark
pair, with a branching ratio of about 58\%~\cite{deFlorian:2016spz}. An
accurate measurement of this channel would allow an extraction of the
Yukawa coupling $y_b$, which is an important
input to Higgs studies. The hadronic environment at the LHC makes this
a challenging prospect -- nevertheless, the decay has been observed
recently by both ATLAS and CMS in the $VH$, or Higgsstrahlung,
production channel~\cite{Sirunyan:2018kst,Aaboud:2018zhk}. QCD
corrections to the partial width are known up to
N\textsuperscript{4}LO~\cite{Baikov:2005rw,Gorishnii:1983cu,Gorishnii:1990zu,
  Gorishnii:1991zr,Becchi:1980vz,Sakai:1980fa,Chetyrkin:1996sr,Chetyrkin:1997vj},
and fully differential NNLO calculations have also been available for
quite some time~\cite{Anastasiou:2011qx,DelDuca:2015zqa}. Recently,
the fully differential N\textsuperscript{3}LO calculation has also
been completed~\cite{Mondini:2019gid}.

The decay channel to gluons, on the other hand, proceeds via a
top-quark loop and contributes around 8\% to the total width. Since
QCD corrections to this channel are indistinguishable from the
$b\bar{b}$-quark case at higher orders in perturbation theory, one
should consider the classes of processes together to obtain a total
hadronic width, as performed in
Ref.~\cite{Davies:2017xsp}. Nevertheless, at NNLO and in the case of
kinematically massless $b$-quarks the processes can be fully
separated, since interference terms between the diagrams
vanish. Related complications which arise at N\textsuperscript{3}LO
have been studied in Ref.~\cite{Mondini:2020uyy}. In the case of
massive $b$-quarks, these interference terms can no longer be
neglected -- their impact has been studied in
Ref.~\cite{Primo:2018zby} and a full NNLO calculation in the massive
case has been carried out in
Refs.~\cite{Bernreuther:2018ynm,Behring:2020uzq}.

In the limit that $M_H\ll 2m_t$, the top-quark loop which couples the
Higgs boson to gluons can be integrated out to obtain an effective
theory with five light active flavours in which the interaction is
local. This simplifies the inclusion of QCD corrections and has
allowed calculations to be performed at
NNLO~\cite{Chetyrkin:1997iv,Caola:2019pfz} and, for the total width,
at N\textsuperscript{3}LO~\cite{Baikov:2006ch} and
N\textsuperscript{4}LO~\cite{Herzog:2017dtz}. The effect of including
a finite top-quark mass on the total width has also been studied in
Ref.~\cite{Schreck:2007um}.

In light of the importance of Higgs physics, several other predictions
at various accuracies and using different approximations are available
beyond those listed here. A complete review of the state of the
theoretical calculations for Higgs boson production and
decay processes, including the calculation of electroweak corrections,
can be found in~Ref.~\cite{Spira:2016ztx}.

In a recent publication~\cite{Gao:2019mlt}, the distributions of the
thrust variable in these decay processes were considered and fixed-order
computations up to approximate NNLO (which contribute at $\ord{\as^3}$
 relative to a Born $H\rightarrow b\bar{b}/gg$ process) were
performed. In that work, the authors noted the poor convergence of the
perturbative series for both processes and were able to show that the
approximate NNLO corrections obtained from the singular terms of a
SCET-derived factorisation formula could ameliorate the scale
dependence of the calculations. They also acknowledged several
shortcomings of their calculation, one of which related to the size of
the logarithms $\log^n \tau/\tau$  which are not resummed
in a purely fixed-order computation and spoil
predictivity in the small $\tau$ region.
Here, we provide resummed predictions at NNLL$'$ accuracy
which complement the results of Ref.~\cite{Gao:2019mlt}.
Compared to NNLL, the resummation at NNLL$'$ accuracy 
incorporates the complete $\ord{\as^2}$
 singular structure for $\Tau_2 \to 0$,{ \it i.e.}
all 2-loop virtual and corresponding real corrections, allowing us to consistently match to NNLO.

Using the
\geneva formalism developed in
Refs.~\cite{Alioli:2012fc,Alioli:2013hqa,Alioli:2015toa}, we are also
able to construct IR-finite events which combine the advantages of the
resummed and fixed-order calculations and are matched to a parton
shower.\footnote{In the gluonic case, resummed predictions for the
  thrust distribution were presented in Ref.~\cite{Mo:2017gzp} and
  compared to parton shower results.}

Having a \geneva implementation of the $H\rightarrow b\bar{b}$ process
will also allow us to produce an NNLOPS generator for the
signal process $pp\rightarrow V(H\to b\bar{b})$. This can be achieved by
combining the Higgs boson decay presented here with our previous
calculation of the $VH$ production process~\cite{Alioli:2019qzz} in
the narrow width approximation. 
Fixed-order calculations for the full
$pp\rightarrow Vb\bar{b}$ process  were performed in
Ref.~\cite{Ferrera:2017zex,Caola:2017xuq,Gauld:2019yng} in the
massless approximation -- more recently, a calculation with massive
$b$-quarks also appeared~\cite{Behring:2020uzq}. An NNLOPS generator
for $W(H\to b\bar{b})$ production via the \minlo method~\cite{Hamilton:2012rf,Hamilton:2013fea,Monni:2019whf,Monni:2020nks}
was presented in~\cite{Astill:2018ivh}, while a separate NNLOPS $H\rightarrow
b\bar{b}$ generator was also made available in
Ref.~\cite{Bizon:2019tfo}. Nonetheless, we believe an independent
implementation of the combined corrections to both production and decay in the \geneva framework will provide a useful
cross-check of previous results. We leave this development to a future
publication.

This paper is organised as follows. In \sec{TheoreticalFramework}, we
briefly explain how resummed predictions are obtained from a
factorisation formula derived in soft-collinear effective theory and
provide numerical results for the resummed $\Tau_2$ distribution in
$H\rightarrow b\bar{b}$ and $H\rightarrow gg$.  In \sec{Genevapart},
we briefly recap the main features of the \geneva method relevant for
the processes at hand. In particular, we discuss various
implementation details, as well as how the matching to the parton
shower is achieved. We present our \geneva results in
\sec{genevaresults}.  Finally, we report our conclusions and
directions for future work in \sec{conc}, while we detail the
construction of the phase space mappings used and the analytical NNLO
decay rates in \apps{tau2map}{NNLOdecrates} respectively.

\section{Resummation from Soft-Collinear Effective Theory}
\label{sec:TheoreticalFramework}
In this section we present, for the first time, the NNLL$'$ resummation of the
 \twoj  observable, $\Tau_2$, for  the decay of a Higgs boson into a pair of  $b$-quarks. 
We also provide results at the same accuracy for the Higgs boson decay into gluons, which were first presented
 in Ref.~\cite{Mo:2017gzp}. We present numerical results for the dimensionless  $\tau \equiv \Tau_2/(2\,M_H)$ distribution,
 where $M_H$ is the mass of the Higgs boson.
\subsection{Formulation}
\label{subsec:form}
Our basic resolution parameter for the hadronic decays of the Higgs
boson is the \twoj, defined as
\begin{align} \label{eq:Tau2def} \Tau_2 = 2 \min_{\vec{n}} \sum_k
  \left( E_k - \lvert\vec{n}\cdot \vec{p_k}\rvert
  \right)\,,\end{align} where $k$ runs over all final state particles
with momenta $p_k = (E_k, \vec{p}_k)$ and $\vec{n}$ is the unit $3$-vector resulting from the minimisation
procedure. In the case where all final-state particles are massless,
it is related to the more familiar thrust $T$, which was widely
studied for $e^+ e^-$ collisions \cite{Catani:1991kz,Catani:1992ua}
and extended to hadronic collisions in Ref.~\cite {Banfi:2004nk}, by
the relation $\Tau_2=2E_{\mathrm{cm}}(1-T)$. For the decays that we
consider, we work in the rest frame of the Higgs boson and always have that
$E_{\mathrm{cm}}=M_H$. Exactly like the thrust variable, the \twoj is
constrained kinematically $(0\leq \Tau_2 \leq M_H)$ and its value is
related to the spatial distribution of the radiation: in the limit
$\Tau_2 \rightarrow 0$ the final state consists of two pencil-like
jets, while for $\Tau_2 \sim M_H$ there are three or more jets
distributed in a more spherical configuration.

We consider the decay rate differential in $\Tau_2$ of a Higgs boson
to either a $b\bar{b}$-quark pair or to a pair of gluons. We consider
massless $b$-quarks with a finite Yukawa coupling to the Higgs
$y_b$. In the gluon case, we work in an effective theory
in which the top-quark loop that couples the Higgs boson
to gluons has been integrated out, leaving an effective local operator
$Hgg$.

  The Born level decay rates for the two processes considered are
  given by
\begin{align}
  \Gamma_\mathrm{B}^b(\mu)=\frac{y_b^2(\mu)\, N_c\, M_H}{8\pi},\qquad
  \Gamma_\mathrm{B}^g(\mu)=\frac{\as^2(\mu)\, G_F\, M_H^3}{36\pi^3\, \sqrt{2}}\,.
\end{align}

It has been shown, both in QCD and SCET, that the differential decay rate factorises in the small $\Tau_2$ limit~\cite{Korchemsky:1999kt,Catani:1992ua,Fleming:2007qr,Schwartz:2007ib} as
\begin{align} \label{eq:facform}
\frac{\df \Gamma^i}{\df \Tau_2} = \Gamma_\mathrm{B}^i(\mu)\ H^i(M_H,\mu)
\int \df p_n^2 \df p_{\bar{n}}^2 \df k \,\delta\left(\Tau_2 -
\frac{2(p_n^2 + p_{\bar{n}}^2)}{M_H} -
k\right)J_n^i(p_n^2,\mu)J_{\bar{n}}^i(p_{\bar{n}}^2,\mu)S^i(k,\mu)
\,\end{align} where the index $i=b,g$ indicates the process in
question.

The decay rate has been factorised into a hard contribution
 $H^i(M_H,\mu)$, a soft function $S^i(k,\mu)$ and two jet functions
 $J_n^i(p_n^2,\mu)$ and $J_{\bar{n}}^i(p_{\bar{n}}^2,\mu)$. The hard
 function is defined as the square of the Wilson coefficients which
 match the full theory (the SM) onto SCET. In the gluon case, an
 additional matching is required from the heavy-top limit effective theory we are working in onto the SM -- thus, the
 hard functions can be written as
  \begin{align} \label{eq:hardfuncs}
   H^b(M_H,\mu) &= |C^b_{\mathrm{SCET}}(M_H,\mu)|^2 \nonumber \\
   H^g(M_H,\mu) &= |C^g_{\mathrm{SCET}}(M_H,\mu)|^2\, |C_t(m_t,\mu)|^2\,.
 \end{align}

 The jet functions describe collinear radiation from the
 Born-level partons along the jet directions $n$ and $\bar{n}$, which
 can be chosen without loss of generality to be orientated along
 $\hat{z}$, viz.~$n=(1,0,0,1)$ and $\bar{n}=(1,0,0,-1)$. The soft
 function accounts for all soft radiation.

Each component in the factorisation theorem must be evaluated at its
own characteristic scale in order to prevent the appearance of large logarithms,
viz.~$\mu_H\sim M_H$, $ \mu_J \sim \sqrt{\phantom{\big(}\!\!\! \Tau_2
  M_H}$, $ \mu_S\sim \Tau_2$. However, since the decay rate must be
evaluated at a single scale, we evolve the separate functions to a
common scale $\mu$ via renormalisation group (RG)
evolution and in so doing resum the large logarithms of
ratios of scales which appear.
  
The resummed spectrum, differential in
the Born kinematics, can then be written as
\begin{align} \label{eq:resummedspectrum}
\frac{\df \Gamma^{i,\rm resum}}{\df \Phi_2 \df \Tau_2}
  &= \frac{\df \Gamma_\mathrm{B}^i(\mu)}{\df \Phi_2} H^i(M_H, \mu_H)\, U^i_H(\mu_H, \mu)
    \otimes \bigl[ S^i(\mu_S) \otimes U^i_S(\mu_S, \mu) \bigr]
\nn\\
& \qquad
\otimes \bigl[ J^i_n (\mu_J) \otimes U^i_J(\mu_J, \mu) \bigr]
\otimes \bigl[ J^i_{\bar{n}} (\mu_J) \otimes U^i_J(\mu_J, \mu) \bigr]
\,\end{align}
where $\df \Gamma_\mathrm{B}^i(\mu) / \df \Phi_2$ indicates the Born
decay rate differential in the two-body phase space and we have used
$\otimes$ to denote the convolutions, dropping the explicit dependence
on the convolution variables. The ingredients necessary for NNLL$'$ accuracy
are all available in the literature and many have been
compiled in Ref.~\cite{Gao:2019mlt}. For the $H\rightarrow b\bar{b}$
case, we take the hard function from Ref.~\cite{Gao:2019mlt}. Since we
are considering only massless $b$-quarks, we can use the soft and jet
functions as implemented in Ref.~\cite{Alioli:2012fc} (and first calculated
at NNLO in Refs.~\cite{Kelley:2011ng,Monni:2011gb,Becher:2006qw})
for $e^+ e^- \to jj$, also recycling the evolution kernels from that work. For the
$H\rightarrow gg$ case, the fully expanded hard function, including
contributions to the $Hgg$ effective vertex from both SCET and the
effective theory where the top-quark is integrated out, appears in
Ref.~\cite{Mo:2017gzp}. The NNLO jet function~\cite{Becher:2010pd} and the evolution kernels are
also given therein, while we obtain the soft function via a Casimir
rescaling of the $H\to b\bar{b}$ case.

\subsection{Numerical results}
\label{subsec:resumresults}
We have implemented the resummed calculation, \eq{resummedspectrum},
in the \geneva framework  (see \sec{Genevapart}) up to NNLL$'$ order, for which we now present results.
Throughout the calculation we choose a Higgs
boson mass $M_H=125.09~\GeV$ and set $\as(M_Z)=0.118$,
$G_F=1.16639\times10^{-5}$. We set the renormalisation scale to be
$\mu_R=M_H$ and run the $b$-quark Yukawa
$y_b$ at two-loop order, setting $m_b=4.92 \GeV$.

\begin{figure}[t]
  \centering
  \includegraphics[width=0.45\textwidth]{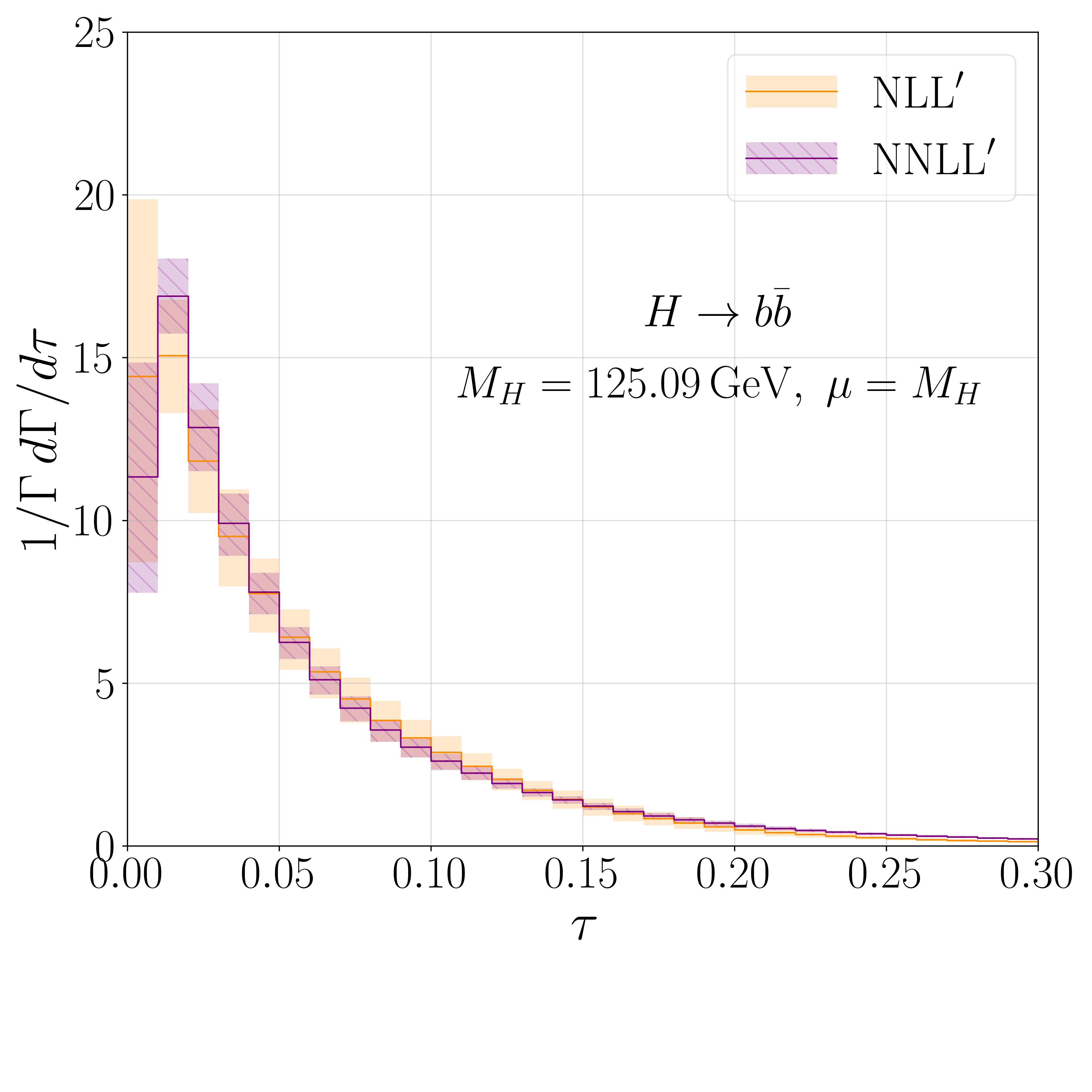}
  \includegraphics[width=0.45\textwidth]{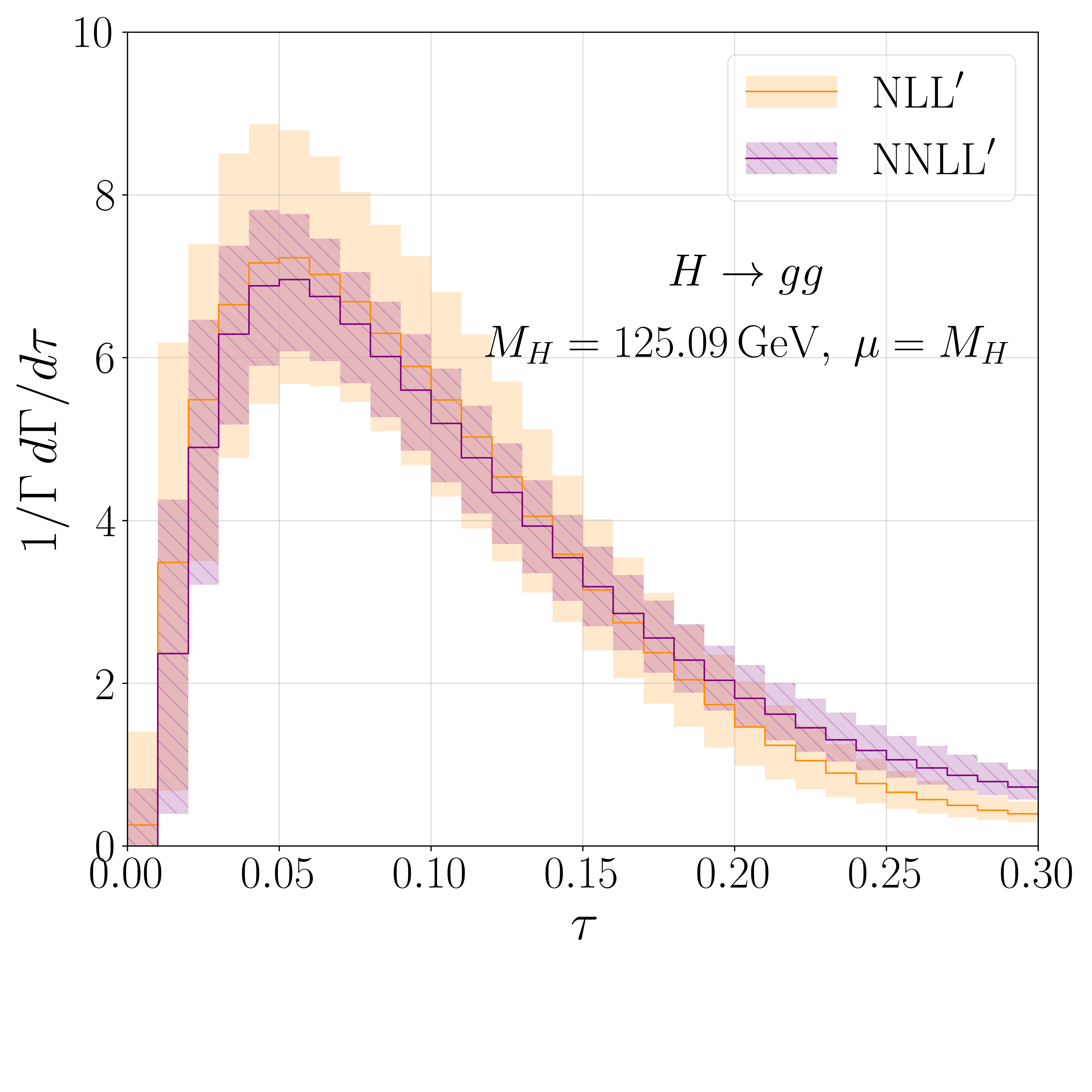}
  \caption{The normalised, resummed spectrum in $\tau \equiv
    \Tau_2/(2M_H)$, corresponding to \eq{resummedspectrum}, at NLL$'$
    and NNLL$'$. The left panel shows the process $H\rightarrow
    b\bar{b}$, the right shows $H\rightarrow gg$.}
  \label{fig:resumonly}
\end{figure}

The normalised distributions in $\tau \equiv \Tau_2/(2M_H)$ are shown
in \fig{resumonly} for both the $H\to b\bar{b}$ and the $H\to gg$
decay channels.
These predictions have been obtained with the scale choices
detailed in \subsec{profscales} and the uncertainties calculated as explained there.

Because of the resummation of large logarithms of $\tau$, our results provide a
physical description at small $\tau$, as can be seen in
\fig{resumonly}, improving on the resummed-expanded results presented
in Ref.~\cite{Gao:2019mlt}.  However, the range of validity of the
resummed calculation is still limited to low $\tau$ values, since the
factorisation formula we rely on (\eq{facform}) is only valid there.
Comparing the two decay channels, we see that the $H \to b\bar b$
process presents a higher peak, located at a lower value of $\tau$,
with a narrower width. The peak of the $\tau$ distribution in the $H\to
gg$ case is instead lower and shifted to larger values of $\tau$, with
a broader width. This behaviour is in line with expectations based on
a naive analysis of the Casimir scaling of the two processes.

In order to extend the validity of the calculation to all values of $\tau$,
one needs to match the resummed result to the fixed-order calculation,
which provides the physical behaviour at large $\tau$. Indeed, in this region, the nonsingular
contribution becomes sizeable and exponentiating the singular contribution is
no longer the correct approach.
The matching of fixed-order calculations to resummation has long been established at
the level of the resummed observable: the most straightforward
approach simply adds the results for the resummed and fixed-order
distributions in the $\tau$ variable and then subtracts the expansion
of the resummed result up to the same order included in the
fixed-order result. In this way the calculation is free from doubly
counted contributions up to the given perturbative order and includes all
the higher-order terms properly resummed.

While the approach just outlined works flawlessly for the $\tau$
distribution we are resumming, it is not directly  applicable to the
construction of a fully exclusive event generator.
In the next section,
\sec{Genevapart}, we show how this can be achieved by means of the
\geneva method, allowing us to perform the matching at the fully differential level.\footnote {For the
  particular process at hand, there are no nontrivial distributions at
  leading-order, so, strictly speaking, one could still perform the
  matching at the level of the $\tau$ distribution and generate the
  other variables needed to achieve a fully exclusive generator
  uniformly in the remaining phase space.}

\section{Implementation in the \geneva framework}
\label{sec:Genevapart}
\subsection{\geneva in a nutshell}
\label{subsec:nutshell}
The \geneva framework allows the matching of a resummed to a
fixed-order calculation and thence to parton shower programs such as
\pythia~\cite{Sjostrand:2014zea}. In so doing, it provides theoretical
predictions which are accurate over the whole phase space and which
describe realistic events of high multiplicity. These can then be
hadronised and fed into the analysis routines used by the experimental
collaborations. The method for separating events into different
multiplicity bins and for performing the matching has already been
described thoroughly in Refs.~\cite{Alioli:2012fc,Alioli:2015toa} and
related references.  Therefore, in this context, we limit ourselves to
restating the primary outcomes as applicable to the case of Higgs
boson hadronic decays up to NNLL$'_\Tau$+NNLO$_2$ accuracy. We remind
the reader that in order to achieve a sensible separation between the
exclusive $3$-jet and the inclusive $4$-jet decay rates, we must also
perform the resummation of $\Tau_3^\cut$ at least at leading-log
order, as we do in the following.

Dropping the process label $i$ for ease of notation, the \geneva Monte
Carlo expressions for the exclusive $2$-jet, $3$-jet and the inclusive
$4$-jet rates are given by\footnote{We make a slight abuse of notation
  in order to highlight the dependence of the $\dgamMC_i$ decay rate
  on the resolution parameters. When an argument contains a single
  term, e.g.\ $\Tau_N^\cut$, it means that the corresponding quantity
  has been integrated over up to the value of the argument. An
  argument \mbox{$\Tau_N > \Tau_N^\cut$} implies instead that the
  corresponding decay rate remains differential in the relevant
  resolution variable for values larger than the cutoff.}
\begin{align} \label{eq:2masterful}
\frac{\dgamMC_2}{\df\Phi_2}(\Tau_2^\cut)
&= \frac{\df\Gamma^{\rm NNLL'}}{\df\Phi_2}(\Tau_2^\cut)
- \biggl[\frac{\df\Gamma^{\rm NNLL'}}{\df\Phi_{2}}(\Tau_2^\cut) \biggr]_{\rm NNLO_2}
+ (B_2 + V_2 + W_2)(\Phi_2)
\nn \\ & \quad
+ \int \! \frac{\df \Phi_3}{\df \Phi_2}\, (B_3 + V_3)(\Phi_3)\, \theta[\Tau_2(\Phi_3) < \Tau_2^\cut]
\nn \\ & \quad
+ \int \! \frac{\df \Phi_4}{\df \Phi_2}\, B_4 (\Phi_4)\, \theta[\Tau_2(\Phi_4) < \Tau_2^\cut]
\,,
\end{align}
\begin{align}
\label{eq:3masterful}
\frac{\dgamMC_{3}}{\df\Phi_{3}} (\Tau_2 > \Tau_2^\cut; \Tau_{3}^\cut)
&= \Bigg\{ \frac{\df\Gamma^{\rm NNLL'}}{\df\Phi_2\df\Tau_2}\cP(\Phi_3)
+ (B_3 + V_3^C)(\Phi_3)
- \biggl[\frac{\df\Gamma^{\rm NNLL'}}{\df\Phi_2\df\Tau_2}\cP(\Phi_3)\,\biggr]_{\NLO_3}\Bigg\}
\nn \\ & \quad
\times U_3(\Phi_3, \Tau_3^\cut)\, \theta(\Tau_2 > \Tau_2^\cut)
\nn \\ & \quad
+\int\!\biggl[\frac{\df\Phi_{4}}{\df\Phi_3^\Tau}\,B_{4}(\Phi_4)\, \theta[\Tau_2(\Phi_4) > \Tau_2^\cut]\,\theta(\Tau_{3} < \Tau_3^\cut)
  \nn \\ & \quad \quad \quad \quad
- \frac{\df\Phi_4}{\df \Phi_3^C}\, C_{4}(\Phi_{4})\, \theta(\Tau_2 > \Tau_2^\cut) \biggr]
\nn \\ & \quad
- B_3(\Phi_3)\, U_3^\one(\Phi_3, \Tau_3^\cut)\, \theta(\Tau_2 > \Tau_2^\cut)
\,,
\end{align}
\begin{align}
\label{eq:3NS}
\frac{\dgamMC_{3}}{\df\Phi_{3}}(\Tau_2 \leq \Tau_2^\cut) &= \bar{\Theta}^{\mathrm{FKS}}_{\mathrm{map}}(\Phi_3)  \, (B_3+V_3)\, (\Phi_3)\,\theta(\Tau_2<\Tau^{\mathrm{cut}}_2) \, ,
\end{align}
\begin{align}
\label{eq:4masterful}
& \frac{\dgamMC_{\geq 4}}{\df\Phi_{4}} (\Tau_2 > \Tau_2^\cut,
\Tau_{3}>\Tau_{3}^\cut) =
\nn \\  & \qquad
\Bigg\{ \frac{\df\Gamma^{\rm NNLL'}}{\df\Phi_2\df\Tau_2}\cP(\Phi_3)
+ (B_3 + V_3^C)(\Phi_3)
- \biggl[\frac{\df\Gamma^{\rm NNLL'}}{\df\Phi_2\df\Tau_2}\cP(\Phi_3)\,\biggr]_{\NLO_3}\Bigg\}
\nn \\ & \qquad
\times U_3'(\Phi_3, \Tau_3)\, \theta(\Tau_2 > \Tau_2^\cut) \Big\vert_{\Phi_3 = \Phi_3^\Tau(\Phi_4)} \!\! \cP(\Phi_4) \, \theta(\Tau_3 > \Tau_3^\cut)
\nn \\ & \qquad
+\bigl\{ B_4(\Phi_4)\,[1 - \Theta^\Tau(\Phi_4)\,\theta(\Tau_3 < \Tau_3^\cut)]
\nn \\ & \qquad
- B_3(\Phi_3^\Tau)\,U_3^{\one\prime}(\Phi_3^\Tau, \Tau_3)\,\cP(\Phi_4)\, \theta(\Tau_3 > \Tau_3^\cut)
\bigr\}\, \theta[\Tau_2(\Phi_4) > \Tau_2^\cut]
\,,\end{align}

\begin{align}
\label{eq:4NS}
\frac{\dgamMC_{\geq 4}}{\df\Phi_{4}}  (\Tau_2 > \Tau_2^\cut, \Tau_{3} \le \Tau_{3}^\cut)
\, = \, &  B_4(\Phi_4)\, \bar{\Theta}^\Tau(\Phi_4)\,\theta(\Tau_3 < \Tau_3^\cut)\, \theta\left(\Tau_2(\Phi_4) > \Tau_2^\cut\right) \, ,
\end{align}
where the $B_j$, $V_j$ and $W_j$ are the $0$-, $1$- and $2$-loop
matrix elements for $j$ partons in the final state.

In the
  equations above, we have introduced the shorthand notation
\begin{align}
\label{eq:dPhiRatio}
 \frac{\df \Phi_{M}}{\df \Phi_N^{\cal O}}  = \df \Phi_{M} \, \delta[ \Phi_N - \Phi^{\cal O}_N(\Phi_M) ] \,\Theta^{\cal O}(\Phi_N)
\,,\end{align}
to indicate that the integration over a region of the $M$-body phase space
is done keeping the $N$-body phase space and the value of some specific
observable $\cal O$ fixed, with $N \leq M$. The $\Theta^{\cal
  O}(\Phi_N)$ term in the previous equation limits the integration to
the phase space points included in the singular contribution for the
given observable $\cal O$.  For example, when generating $3$-body
events we use
\begin{equation} \label{eq:Phi3TauProj}
\frac{\df\Phi_4}{\df\Phi_3^\Tau} \equiv \df\Phi_4\,\delta[\Phi_3 - \Phi^\Tau_3(\Phi_4)]\,\Theta^\Tau(\Phi_4)
\,,\end{equation}
where the map used by the $3 \to 4$ splitting has been constructed
to preserve $\Tau_2$, i.e.
\begin{equation} \label{eq:Tau2map}
\Tau_2(\Phi_3^\Tau(\Phi_4)) = \Tau_2(\Phi_4)
\,\end{equation}
and $\Theta^\Tau(\Phi_4)$ defines the projectable region of $\Phi_4$ which can be reached starting from a point in $\Phi_3$ with a specific value of $\Tau_2$.
The usage of a $\Tau_2$-preserving mapping is necessary to ensure that
the pointwise singular $\Tau_2$ dependence is alike among all terms in
\eqs{3masterful}{4masterful} and that the cancellation of said singular
terms is guaranteed on an event-by-event basis.

The expressions in \eqs{3NS}{4NS} encode the nonsingular contributions
to the $3$- and $4$-jet rates which arise from non-projectable
configurations below the corresponding cut. This is highlighted by the
appearance of the complementary $\Theta$ functions, $\bar{\Theta}^{\cal O}$, which account for
any configuration which is not projectable either because it would
result in an invalid underlying-Born flavour structure or because it
does not satisfy the $\Tau_2$-preserving mapping (see also
Ref.~\cite{Alioli:2019qzz}).

The term $V_3^C$ denotes the soft-virtual contribution of a standard NLO local subtraction (in our implementation, we follow
the FKS subtraction as detailed in Ref.~\cite{Frixione:2007vw}). We have that
\begin{align} \label{eq:FOFKS}
  V_3^C(\Phi_3) = V_3(\Phi_3)+\int\frac{\df\Phi_4}{\df \Phi_3^C}C_4(\Phi_4)\,,
\end{align}
 with $C_4$ a singular approximation of $B_4$: in practice we use the subtraction counterterms which we integrate over the radiation variables  $\df\Phi_4 / \df \Phi_3^C$  using the singular limit $C$ of the phase space mapping. $U_3$ is a LL Sudakov which resums large logarithms of $\Tau_3$ and $U_3'$ its derivative with respect to $\Tau_3$. Its exact form is given by
\begin{align}
\label{eq:tau3sudakov}
U_3(\Tau_3^{\mathrm{cut}},\Tau_3^{\mathrm{max}}) = \exp\left[-\frac{\as}{2\pi}C_k\log^2\left(\frac{\Tau_3^{\mathrm{max}}}{\Tau_3^{\mathrm{cut}}}\right)\right]
\end{align}
where the Casimir factor $C_k$ depends on the flavour content of the $3$-jet
event,\\ \mbox{$C_{q\bar{q}g}=2C_F+C_A$} or \mbox{$C_{ggg}=3C_A$}, and we run the coupling at NNLL order.

The term $\mathcal{P}(\Phi_{N+1})$
represents a normalised splitting probability which
serves to extend the differential dependence of the resummed terms
from the $N$-jet to the $(N\!+\!1)$-jet phase space. For example, in
\eq{3masterful}, the term $\cP(\Phi_3)$ makes the resummed spectrum in
the first term (which is naturally differential in the $\Phi_2$
variables and $\Tau_2$) differential also in the additional two
variables needed to cover the full $\Phi_3$ phase space. These
splitting probabilities are normalised, i.e. they satisfy
\begin{align}
\label{eq:cPnorm}
\int \! \frac{\df\Phi_{N+1}}{\df \Phi_{N} \df \Tau_N} \, \cP(\Phi_{N+1}) = 1
\,.\end{align}
The two extra variables are chosen to be an energy ratio $z$ and an
azimuthal angle $\phi$. In the soft and collinear limit,
$z=E_{\mathrm{sister}}/(E_{\mathrm{sister}}+E_{\mathrm{daughter}})$
where the daughter and the sister are assigned to be the pair of
particles that are closest according to the \nj metric and which
therefore set the value of $\Tau_N$, i.e. which minimise the quantity
\begin{equation}
    \rho_{ij}=|\vec{p}_i|+|\vec{p}_j|-|\vec{p}_i+\vec{p}_j|.
\end{equation}
The daughter particle is defined
to be the gluon for $q\rightarrow qg$ splittings, the quark for
$g\rightarrow q\bar{q}$ and the softer gluon for $g\rightarrow gg$.
These definitions in hand, the normalised splitting probability is
given by
\begin{align}
\label{eq:fullP}
  \cP(\Phi_{N+1})=\frac{\mathrm{AP}_{\mathrm{sp}}(z,\phi)}{\sum_{\mathrm{sp}}\int^{z_\mathrm{max}(\Tau_N)}_{z_\mathrm{min}(\Tau_N)}\df z\df\phi \mathrm{AP}_{\mathrm{sp}}(z,\phi)}\frac{\df \Phi_N \df \Tau_N \df z\df \phi}{\df \Phi_{N+1}}
  \,,
\end{align}
where $\mathrm{AP}_{\mathrm{sp}}(z,\phi)$ is the unregularised
Altarelli-Parisi splitting function.\footnote{We note that the
  discussion here is simplified in the case of the Higgs
  boson decay as only final-state radiation is present. A
  more detailed discussion including the case of initial-state
  radiation may be found in Ref.~\cite{Alioli:2015toa}.}

The implementation of the splitting probability requires us to
construct the full $\Phi_{N+1}$ phase space from $\Phi_N$ and a
value of $\Tau_N$. Similarly, we mentioned above that the real
integration in the fixed-order part of the calculation requires us to
project from $\Phi_{N+1}$ configurations onto $\Phi_N$ while
preserving the value of $\Tau_2$. Both of these tasks demand a map
that satisfies \eq{Tau2map} -- the construction of such a map is
detailed in \app{tau2map}.

\subsection{Implementation details}
\label{subsec:genevadetails}
In this section we discuss the particulars of the implementation of
the Higgs boson decay processes in \geneva. Throughout
this section we use the same settings and values for SM parameters as
in \subsec{resumresults}. In the $H\rightarrow b\bar{b}$ case, we
implement the analytic matrix elements found in
Ref.~\cite{DelDuca:2015zqa}, while in the $H\rightarrow gg$ case we
interface to the \textsc{OpenLoops} package~\cite{Cascioli:2011va,Buccioni:2017yxi,Buccioni:2019sur}.

\subsubsection{Profile scales}
\label{subsec:profscales}

The resummation provided by the RGE of the functions in \eq{facform}
correctly accounts for logarithms of the form $\log(\Tau_2/2M_H)$
which become large in size for small values of $\Tau_2$. In the
fixed-order region, however, where $\Tau_2$ is larger, such logarithms
are more modest in size and continuing to resum them would introduce
undesirable higher-order contributions.

We must therefore switch off the resummation before this happens. This
can be achieved by setting all scales to a common nonsingular scale in
the fixed-order region, $\mu_{\mathrm{NS}}=\mu_S=\mu_J=\mu_H$, which
stops the evolution ensuring that the resummed contribution is
cancelled out exactly by the resummed-expanded.  In order to achieve a
smooth transition between the resummation and the fixed-order (FO)
regimes, we make use of profile scales $\mu_J(\Tau_2)$ and
$\mu_S(\Tau_2)$ which interpolate between the characteristic scales  and
$\mu_{\mathrm{NS}}$~\cite{Abbate:2010xh,Berger:2010xi}. Specifically,
we have that
\begin{align}
  \mu_H&=\mu_{\mathrm{NS}}\nn\\
  \mu_S(\Tau_2)&=\mu_{\mathrm{NS}}f_{\mathrm{run}}(\Tau_2/2M_H)\nn\\
  \mu_J(\Tau_2)&=\mu_{\mathrm{NS}}\sqrt{f_{\mathrm{run}}(\Tau_2/2M_H)}\,,
  \label{eq:jetsoftscales}
\end{align}
where the common profile function $f_{\mathrm{run}}(x)$ is given by
\begin{align}
f_{\rm run}(x) &=
\begin{cases} \vspace{1mm} \displaystyle ax_0 \left[1+ \left(\frac{x}{2x_0}\right)^2 \right] & x \le 2x_0\,,
 \\ \vspace{1mm} ax & 2x_0 \le x \le x_1\,,
 \\ \vspace{1mm} \displaystyle ax + \frac{(2-ax_2-ax_3)(x-x_1)^2}{2(x_2-x_1)(x_3-x_1)} & x_1 \le x \le x_2\,,
 \\ \vspace{1mm} \displaystyle 1 - \frac{(2-ax_1-ax_2)(x-x_3)^2}{2(x_3-x_1)(x_3-x_2)} & x_2 \le x \le x_3\,,
 \\ 1 & x_3 \le x\,.
\end{cases}
\label{eq:frun}
\,.\end{align}
This form has strict canonical scaling below $x_1$ and switches off
the resummation above $x_3$; for $a=1$ it matches the form of the
profiles used in e.g.\ Ref.~\cite{Stewart:2013faa}.

\begin{figure}[h!]
  \centering
  \includegraphics[width=0.45\textwidth]{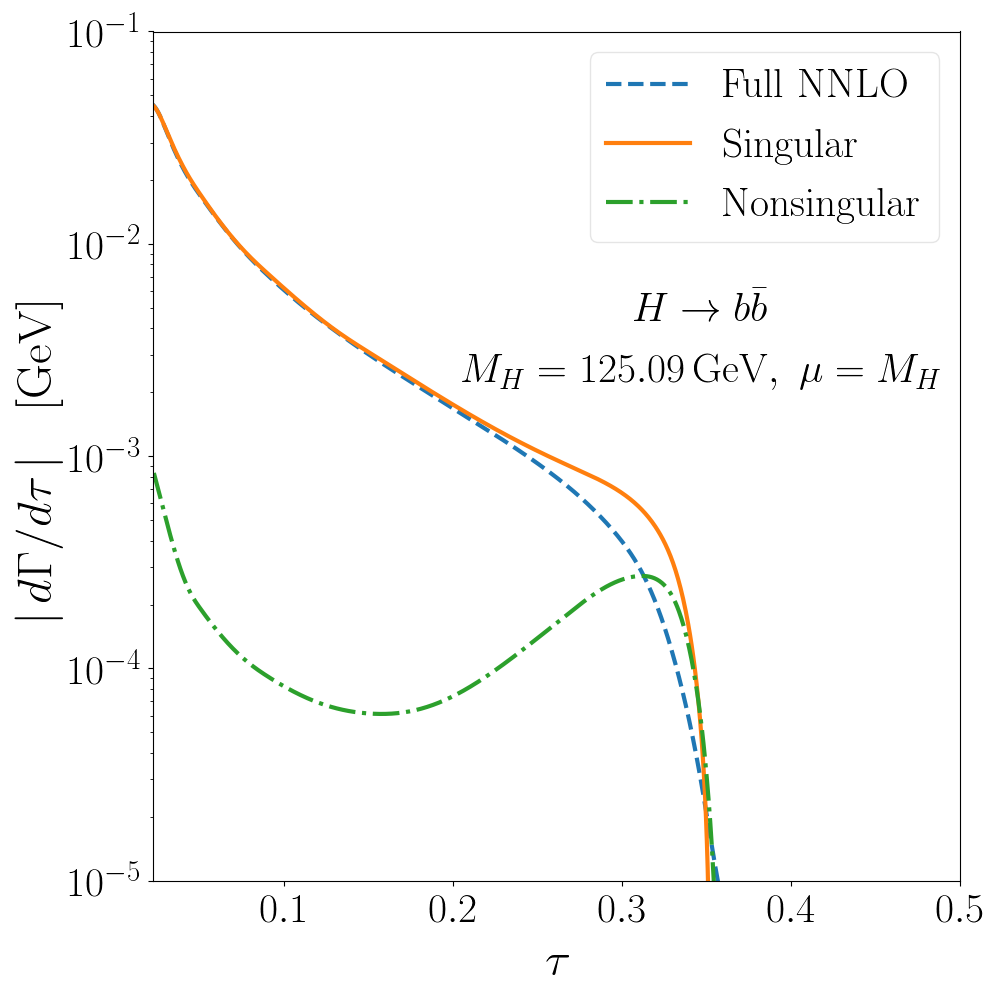}%
  \includegraphics[width=0.45\textwidth]{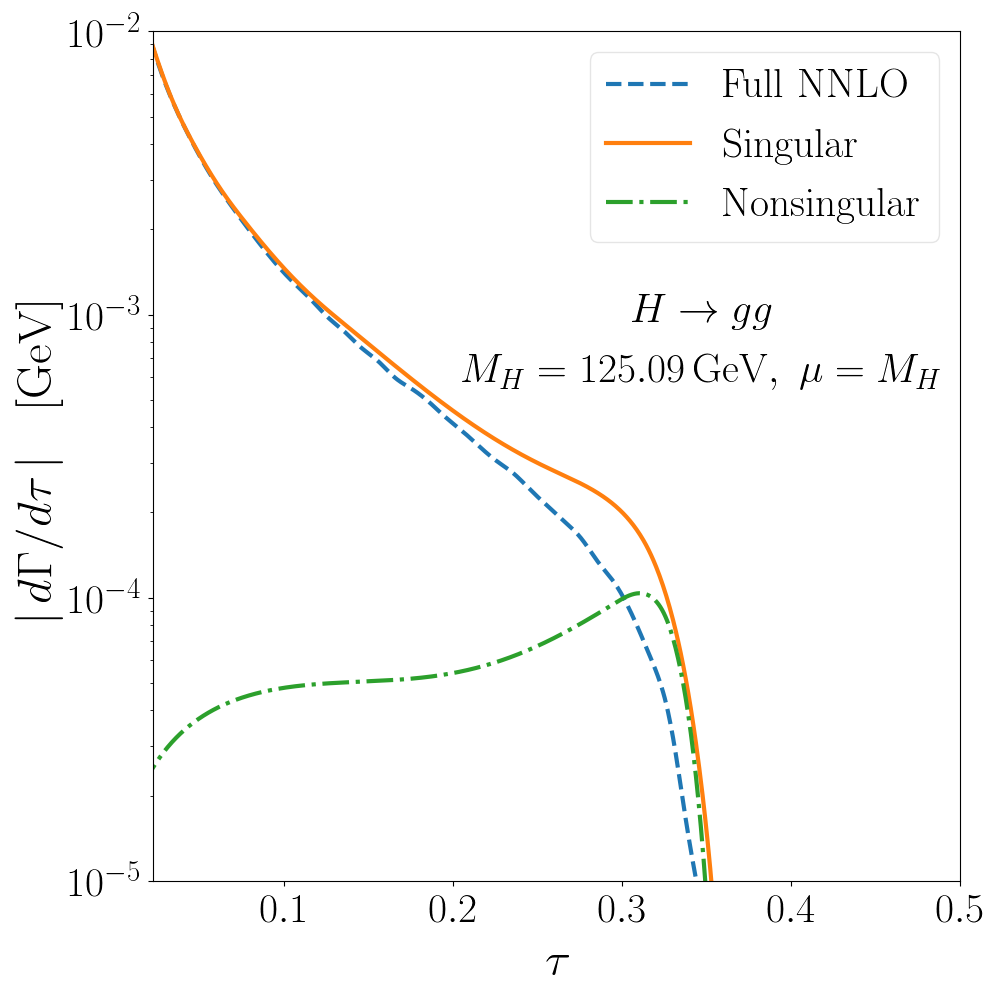}
  \caption{The full, singular and nonsingular contributions to the
    Higgs boson decay rate as a function of $\tau \equiv
    \Tau_2/(2M_H)$. The left panel shows the process $H\rightarrow
    b\bar{b}$, the right shows $H\rightarrow gg$.}
  \label{fig:nonsingplot}
\end{figure}

In order to determine the choice of parameters $a,\,x_i$ it is
instructive to examine the relative sizes of the singular and
nonsingular contributions as a function of $\Tau_2$ to determine where
the resummation should be switched off. This is done for the two
decay channels as shown in \fig{nonsingplot}. We see that the
singular and nonsingular pieces become similar in size at around $\tau
\equiv \Tau_2/(2M_H) \approx 0.3$, and therefore set for both
processes
\begin{align} \label{eq:Tauprofile}
\mu_\FO \equiv \mu_{\mathrm{NS}} &= M_H \,, &\{x_1,x_2, x_3\} &= \{0.1, 0.2, 0.3\}
\,.\end{align}
We notice that in the limit $\tau\rightarrow 0$ the singular
contribution becomes an increasingly good approximation to the
fixed-order result, reflecting the proper cancellation of the singular
terms between the fixed-order and resummed-expanded parts of the
calculation. We set the remaining parameters $a=1,\,ax_0=3\,\GeV/M_H$
for the $gg$ channel following Ref.~\cite{Mo:2017gzp}, while for the
$b\bar{b}$ channel we set $a=1/2,\,ax_0=2.5\,\GeV/M_H$.

\begin{figure}
  \centering
  \includegraphics[width=0.45\textwidth]{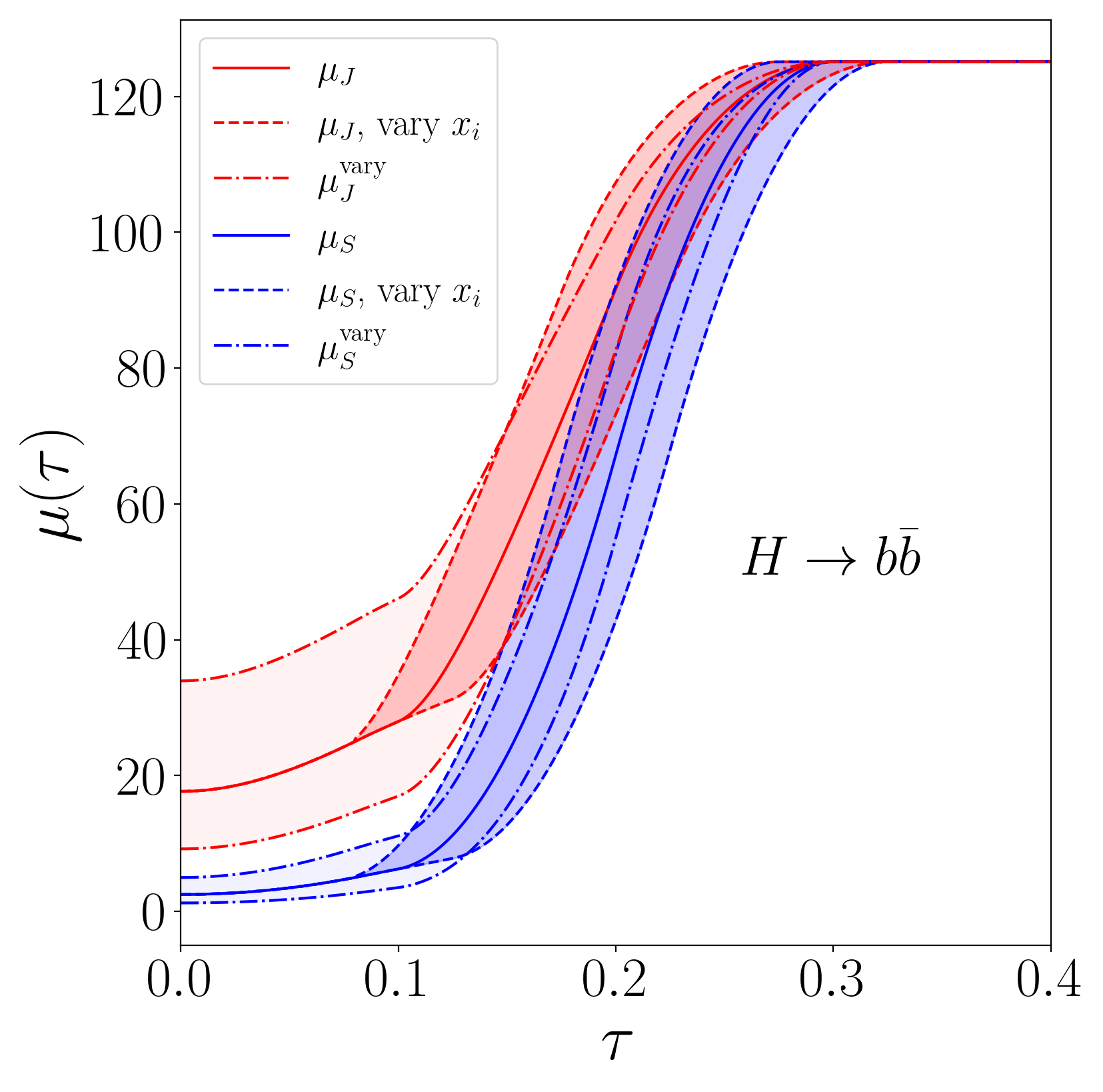}
  \includegraphics[width=0.45\textwidth]{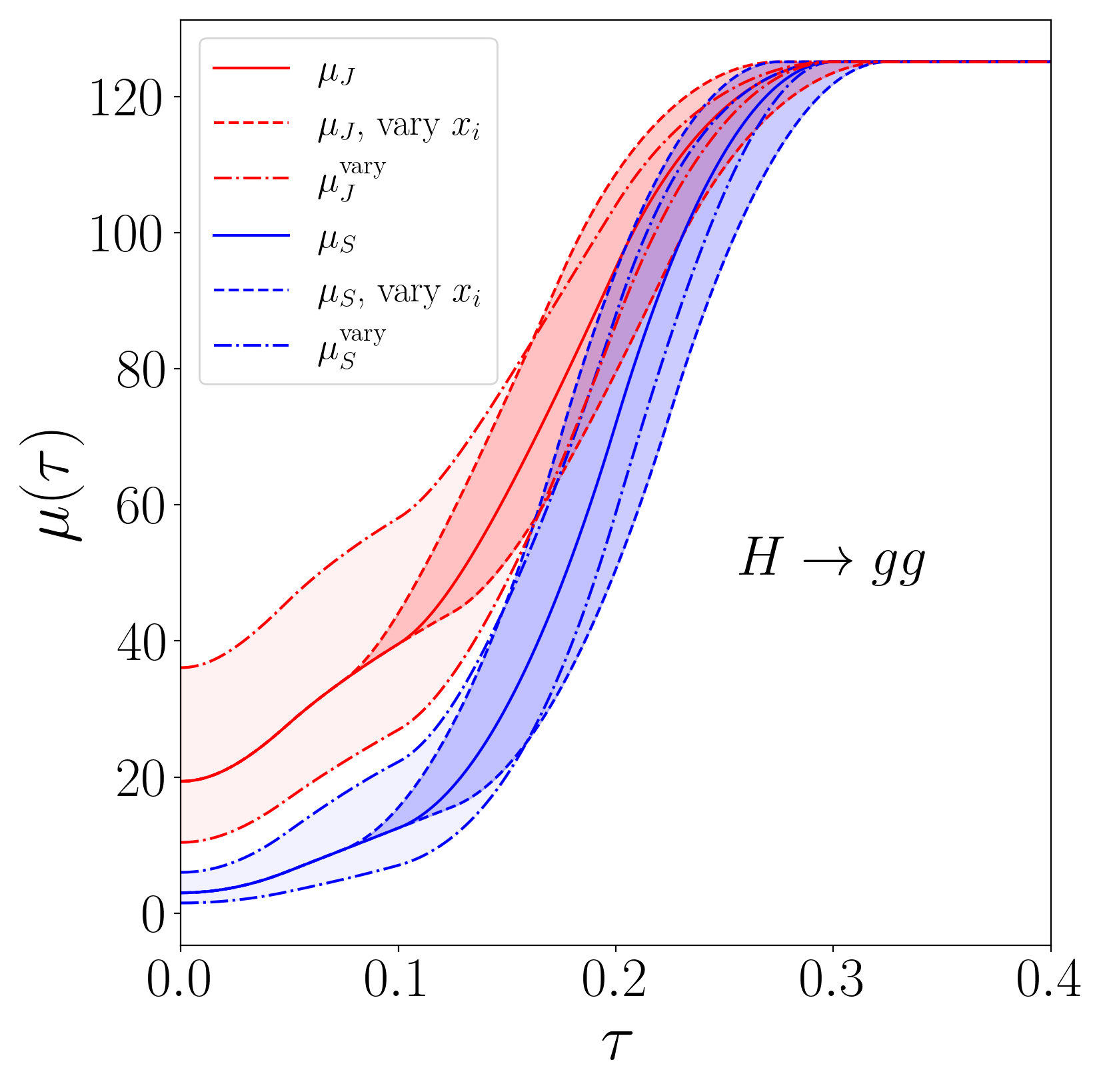}
  \caption{The profile scales from \eq{jetsoftscales} for the $b\bar{b}$ case (left) and $gg$ case (right).
    The red (blue) curves show the jet (soft)
    scale $\mu_J$ ($\mu_S$) and its variations. Variations due to shifts in the transition points $x_i$ are
    shown as dashed lines, while the variations in \eq{fvary} are shown as dashed-dotted lines. Variations
  of $\mu_H$ result in an overall rescaling of these curves and hence are not shown.}
  \label{fig:profscales}
\end{figure}

The uncertainties associated with the resummed and fixed-order
calculations are estimated by varying the profile scales. For the uncertainty arising from the FO part, we adopt the usual prescription of varying $\mu_\FO$ up and down
by a factor of 2 and taking the maximal absolute deviation from the
central value as a measure of the uncertainty. This preserves
everywhere the ratios between the various scales $\mu_H$, $\mu_J$ and
$\mu_S$ and so the arguments of the logarithms which are resummed by
the RGE factors are unaffected. In the resummed case, we vary the
profile scales for $\mu_J$ and $\mu_S$ about their central profiles
while keeping $\mu_H=\mu_\FO$ fixed. Specifically, defining a
variation function (see e.g.\ Ref.~\cite{Gangal:2014qda})
\begin{align}
  f_{\mathrm{vary}}(x)=\begin{cases}
                    2(1-x^2/x_3^2) &  0\leq x \leq x_3/2  \\
                    1+2(1-x/x_3)^2 & x_3/2\leq x\leq x_3  \\
                    1 & x_3\leq x
  \end{cases}\,,
\end{align}
we vary the soft- and jet-function scales such that
\begin{align}
  \mu_S^{\uparrow}(x) &= f_{\mathrm{vary}}(x)\mu_{S}(x)\,,\nn\\
  \mu_S^{\downarrow}(x) &= f_{\mathrm{vary}}(x)^{-1}\mu_{S}(x)\,,\nn\\
  \mu_J^{\uparrow}(x) &= \mu_S(x)^{1/2-\eta}\mu_{\FO}^{1/2+\eta}\,,\nn\\
  \mu_J^{\downarrow}(x) &= \mu_S(x)^{1/2+\eta}\mu_{\FO}^{1/2-\eta}\,,
  \label{eq:fvary}
\end{align}
where $\eta=1/6$. In this way the arguments of the resummed logarithms
are varied in order to estimate the size of higher-order corrections
in the resummed series while maintaining the scale hierarchy
$\mu_\FO\sim \mu_H \gg \mu_J \sim \sqrt{\mu_H\mu_S} \gg \mu_S$. More
details on the specifics of this prescription may be found in
Ref.~\cite{Gangal:2014qda}. In addition, we include two more profiles
where we vary all $x_i$ transition points by $\pm0.025$
simultaneously. We thus obtain 6 profile variations in total and take
the maximal absolute deviation in the result from the central value as
the resummation uncertainty. The total uncertainty is then obtained as
the quadrature sum of the resummation and fixed-order uncertainties.
The profiles and their variations are shown in \fig{profscales} for
both the $b\bar{b}$ and $gg$ cases.

\subsubsection{Comparing the spectrum and the derivative of the cumulant}
Since the profile scales which we just discussed have themselves a
functional dependence on $\Tau_2$, the integral of the spectrum that
one obtains from \eq{resummedspectrum} is not exactly equal to the
cumulant in \eq{2masterful}  evaluated at the highest scale.

Choosing canonical scaling, i.e. $\mu \propto \Tau_2^{\rm cut}$,  we have
\begin{align}
\label{eq:problem}
 &  \int_{0}^{\Tau_2^{\max}} \frac{\df\Gamma^{\NNLL'}}{\df\Phi_2 \df\Tau_2}(\mu(\Tau_2))\df\Tau_2 =  \frac{\df\Gamma^{\NNLL'}}{\df\Phi_2}(\Tau_2^{\max},\mu(\Tau_2^{\max}))+ \mathcal{O}(\rm N^3LL)\,,
\end{align}
where ${\Tau_2^{\max}}$ is the upper kinematical limit. By
integrating the spectrum we therefore obtain not only the cumulant but also
unwanted additional terms of higher order. Depending on the convergence
properties of the perturbative series, these additional terms can be
numerically relevant and cause a sizeable difference between the inclusive
decay rate obtained from the resummed calculation and the FO result.
To obviate this problem, we supplement the spectrum in \eqs{3masterful}{4masterful} with an additional
higher order term. The contribution of this term is restricted to the
region of $\Tau_2$ where the spurious $\rm N^3LL$ terms are sizeable,
and vanishes in the FO region; crucially, upon integration it ensures
that the FO rate is recovered. It takes the form:
\begin{equation}
\label{eq:term}
\kappa(\Tau_2) \left[ \frac{\df}{\df \Tau_2} \frac{\df \Gamma^{\rm NNLL'}}{\df \Phi_2}(\Tau_2, \mu_h(\Tau_2))
 - \frac{\df \Gamma^{\rm NNLL'}}{\df \Phi_2 \df \Tau_2} (\mu_h(\Tau_2)) \right],
\end{equation}
where $\kappa(\Tau_2)$ and $\mu_h(\Tau_2)$ are smooth functions. It is
clear that this vanishes in the FO region where $\mu_h(\Tau_2)\sim
M_H$ as required -- in order to restrict its contribution further, we
also choose $\kappa(\Tau_2)$ to tend to zero in this region to
minimise its size before exact cancellation is reached and choose the
profile scale $\mu_h(\Tau_2)$ to reach $M_H$ at a lower value of
$\Tau_2$ than the rest of the calculation. This prevents the accuracy
of the tail of the spectrum from being spoiled, while keeping the resulting 
changes in  the peak region contained within its scale uncertainty band. We tune
$\kappa(\Tau_2)$ to recover the correct inclusive rate, both for the
central FO scale and also for its variations such that the result of
integration is identical to a FO calculation for inclusive quantities.

\subsubsection{Power-suppressed corrections to the nonsingular cumulant}
\begin{figure}
  \centering
  \includegraphics[width=0.49\textwidth]{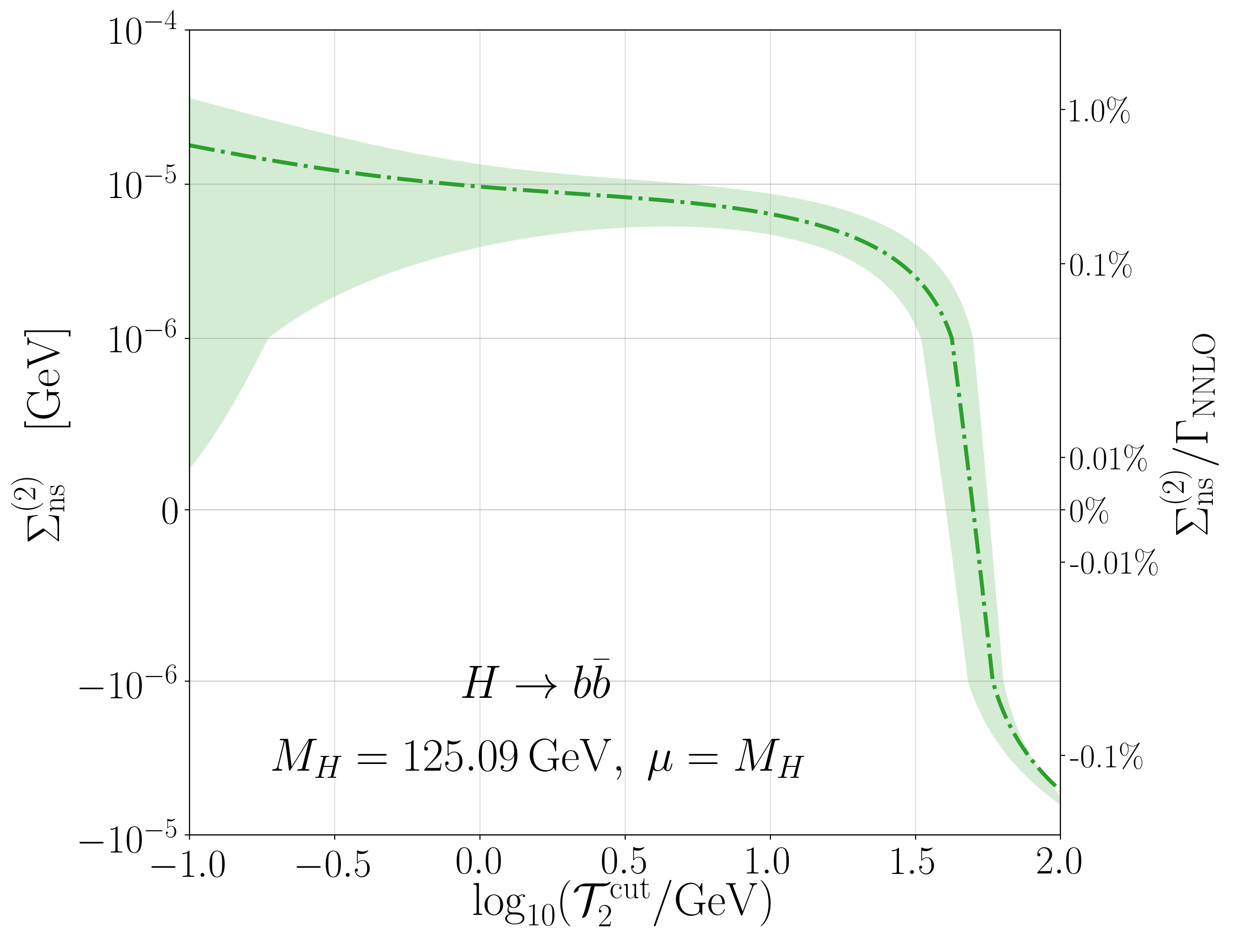}
  \includegraphics[width=0.49\textwidth]{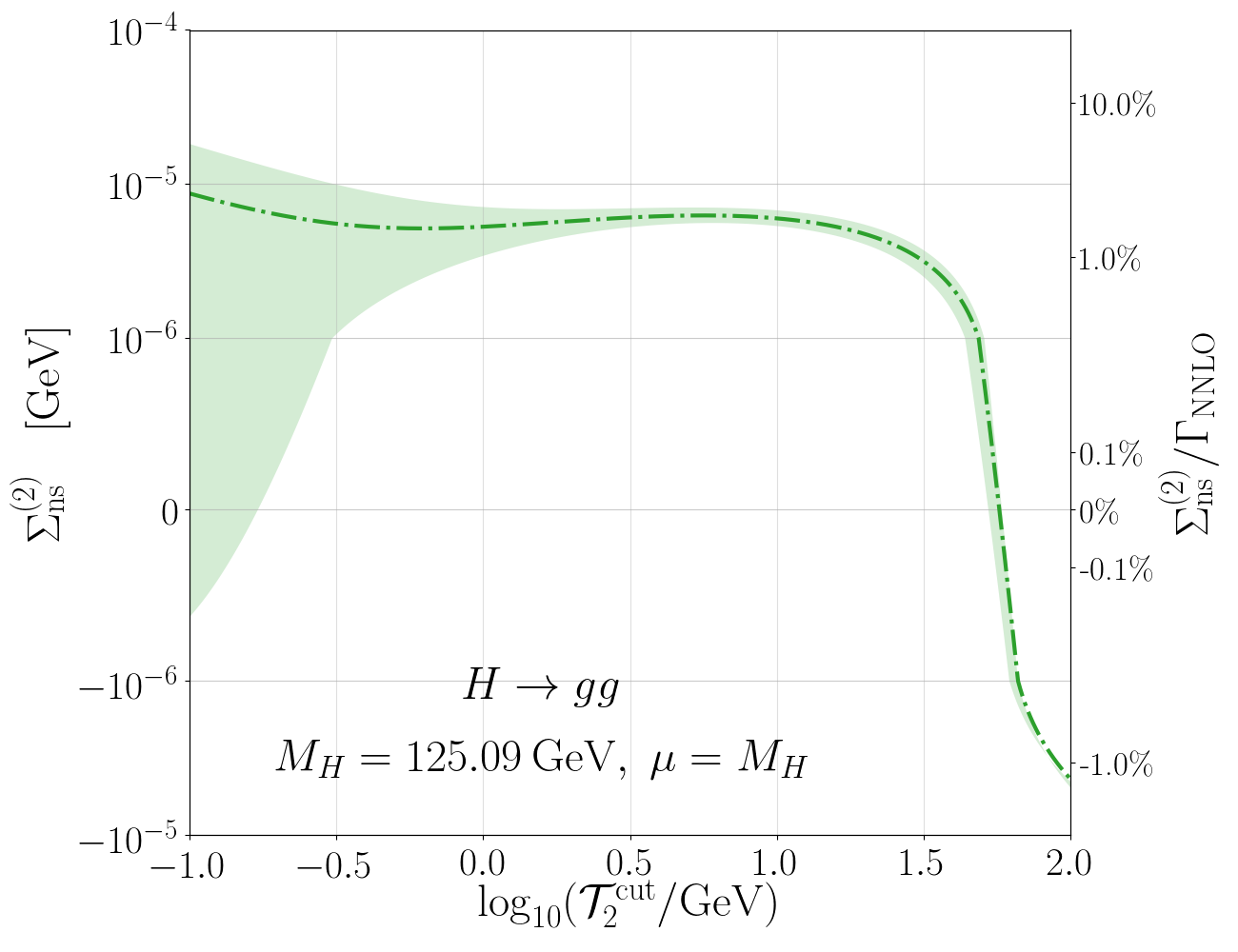}
  \caption{The missing $\ord{\as^2}$ nonsingular contribution to the
    cumulant as a function of $\Tau_2^\cut$ for the $H \to b\bar{b}$
    process (left) and the $H \to gg$ (right). The shaded band denotes
    the statistical error on the spectrum rather than the cumulant,
    which we include to help gauge the size of the uncertainty on the
    power corrections as a function of $\Tau_2^\cut$.}
  \label{fig:nscum}
\end{figure}

The integration of the differential decay rate in \eq{2masterful} over
the $\Phi_2$ phase space produces an NNLO accurate total width. For
differential quantities, however, the $\ord{\as^2}$ terms in
\eq{2masterful} are guaranteed to be NNLO accurate only up to power
corrections in $\Tau_2^\cut$ since any projective map one could devise
could not preserve all $\Phi_2$ quantities simultaneously. This
fundamental limit on the accuracy of event generators actually allows
us to sidestep the problem of implementing a full NNLO subtraction --
since the total width is the only quantity that is certain to be NNLO
accurate, we can drop all the $\ord{\as^2}$ terms in the cumulant and
achieve the correct NNLO width by reweighting. That is, rather than
implementing the full form of \eq{2masterful}, we instead use
\begin{align}
\frac{\widetilde{\dgamMC_2}}{\df\Phi_2}(\Tau_2^\cut)
&= \frac{\df\Gamma^{\rm NNLL'}}{\df\Phi_2}(\Tau_2^\cut)
- \biggl[\frac{\df\Gamma^{\rm NNLL'}}{\df\Phi_{2}}(\Tau_2^\cut) \biggr]_{\rm NLO_2}
\nn + B_2(\Phi_2)  + V_2(\Phi_2)
 \\ & \quad
+ \int \! \frac{\df \Phi_3}{\df \Phi_2}\, B_3 (\Phi_3)\, \theta\left(\Tau_2(\Phi_3) < \Tau_2^\cut\right)
\,,\label{eq:2tilde}
\end{align}
which requires only a local NLO subtraction. The remaining nonsingular
terms take the form
\begin{align} \label{eq:Gammanons}
\frac{\df\Gamma_2^\nons}{\df\Phi_{2}}(\Tau_2^\cut)
&= \bigl[ \as f_1(\Tau_2^\cut, \Phi_2) + \as^2 f_2(\Tau_2^\cut, \Phi_2) \bigr]\Tau_2^\cut
\end{align}
where the functions $f_i(\Tau_2^{\rm cut},\Phi_2)$ are at worst
logarithmically divergent in the small $\Tau_2^{\cut}$ limit. We
include the NLO term proportional to $f_1(\Tau_2^\cut, \Phi_2)$ in
\eq{2tilde} via an on-the-fly $\NLO_2$ calculation, but neglect the
$f_2(\Tau_2^\cut, \Phi_2)$ piece. The size of this neglected term as a
function of the cut is shown in \fig{nscum} for both processes. We see
that at our default value of $\Tau_2^\cut=1\GeV$ the missing
$\ord{\as^2}$ terms are of a size $\sim 10^{-5}~\GeV$ in both cases.
This amounts to a relative correction of  $\ord{0.4 \%}$ for the $b\bar{b}$ channel and of $\ord{1 \%}$ for the $gg$. 
Smaller power corrections could
naturally also be obtained by modifying the factorisation formula
\eq{facform} to include subleading power contributions
\cite{Moult:2019mog,Moult:2018jjd} or by lowering further the value of
$\Tau_2^\cut$. In this limit, however, the calculation suffers from
numerical problems originating from the stability of the matrix
elements and of the NLO subtraction procedure close to extreme soft or
collinear configurations, which motivates our default choice.

In order to correct for this discrepancy and obtain the correct NNLO
inclusive decay width, we may simply rescale the weights of the
$\Phi_2$ events in such a way that we match the known analytic result
at NNLO. We are thus able to include the effects of the $f_2$ term in
\eq{Gammanons} on the total cross section that would have been present
had we implemented \eq{2masterful} literally. Since neither
\eq{2masterful} nor our approach in \eq{2tilde} achieves the exact
$\ord{\as^2}$ $\Phi_2$ dependence of all observables, our
approximation does not inherently limit the accuracy of our
predictions.

\subsubsection{Interface to the parton shower}
We briefly recap the main features of the parton shower interface in
\geneva here and refer the interested reader to section 3 of
Ref.~\cite{Alioli:2015toa} for a more detailed discussion.

The partonic jet decay rates $\dgamMC_2$, $\dgamMC_3$ and
$\dgamMC_{\ge 4}$ each include contributions from higher multiplicity
phase space points, but only in those cases where
$\Tau_N(\Phi_M)<\Tau_N^\cut$. In order to make the calculation fully
differential in the higher multiplicities, a parton shower is
interfaced which adds radiation to each jet decay rate in a unitary
and recursive manner. Ideally, the shower should leave the values of the jet rates and their accuracy unaffected, restoring the
  emissions in $\dgamMC_2$ and $\dgamMC_3$ which were integrated over
  when the jet rates were constructed and also adding extra final-state
  partons to the inclusive $\dgamMC_{\ge 4}$.

For illustrative purposes, we consider a shower strongly ordered in
$\Tau_N$, such that $\Tau_2(\Phi_3)\gg\Tau_3(\Phi_4)\gg\dots$
A shower history of this kind could be constructed by taking the output of a
shower ordered in a more conventional variable and reclustering the
partons using the \nj metric $\Tau_N$.

In general, the requirement of the
preservation of the accuracy of the jet rates after applying the shower
on a phase space point $\Phi_N$ sets constraints on the point
$\Phi_{M}$ reached after the shower.
For the cases in which the showered events originate from $\Phi_2$
events, the main constraint is that the integral of the decay rate
below the $\Tau_2^\cut$ (which is NNLL$'$+NNLO accurate) must not be
modified. The emissions generated by the shower must in this case
satisfy $\Tau_2(\Phi_N)<\Tau_2^\cut$, so that they recover the events
which were integrated over in the construction of the 2-jet exclusive
decay rate and add events with more emissions below the cut. In case
of a single shower emission we require also that the resulting
$\Phi_3$ point is projectable onto $\Phi_2$, as these are the only
configurations at this order which are included in \eq{2tilde}. Both
of these conditions can be implemented with a careful choice of the
starting scale of the shower. The preservation of the decay rate below
the cut is then ensured by the unitarity of the shower evolution. In
practice, we allow for a tiny spillover up to $5\%$ above
$\Tau_2^\cut$ in order to smoothen the transition.

The showering of $\Phi_3$ and $\Phi_4$ events must be treated more
carefully in order to preserve the NNLL$'$+NNLO accuracy of the
$\Tau_2$ spectrum. Crucially, we must ensure that the $\Phi_4$ points
produced after the first emission are projectable onto $\Phi_3$ using
the $\Tau_2$-preserving map discussed in \app{tau2map}. Since the
shower cannot guarantee this, we instead perform the first emission in
\geneva (using the analytic form of the LL Sudakov factor and phase
space maps) and only thereafter allow the shower to act as usual,
subject to the restriction $\Tau_4(\Phi_N)\leq\Tau_3(\Phi_4)$. We apply this
procedure only to the $\Phi_3$ events and find that
\begin{align}
  \frac{\mathrm{d} \Gamma_{3}^{\mathrm{MC}}}{\df\Phi_{3}} (\Tau_2 > \Tau_2^\cut, \Tau_3^\cut, \Lambda_3) &= \frac{\dgamMC_{3}}{\df\Phi_{3}} (\Tau_2 > \Tau_2^\cut, \Tau_3^\cut) \, U_3(\Tau_3^\cut, \Lambda_3)
\end{align}

\begin{align}
\frac{\mathrm{d} \Gamma_{\geq 4}^{\mathrm{MC}}}{\df\Phi_{4}} (\Tau_2 > \Tau_2^\cut, \Tau_3^\cut,\Tau_{3}>\Lambda_3)  = \frac{\dgamMC_{\geq 4}}{\df\Phi_{4}} (\Tau_2 > \Tau_2^\cut, \Tau_{3}>\Tau_{3}^\cut) 
\\\nn
\quad + \frac{\df}{\df \Tau_3} \, \frac{\mathrm{d} \Gamma^{\mathrm{MC}}_{3}}{\df\Phi_{3}} (\Tau_2 > \Tau_2^\cut, \Tau_3^\cut, \Tau_3) \times\cP(\Phi_4) \,  \theta( \Lambda_3<\Tau_3<\Tau_3^{\max})
\,.
\end{align}

By choosing $\Lambda_3\sim \Lambda_{\mathrm{QCD}}$, the Sudakov factor
$U_3(\Tau_3^\cut,\Lambda_3)$ becomes vanishingly small and we can
relax the shower conditions on the $3$-jet contributions. The showered
events therefore originate almost exclusively from either $\dgamMC_2$ or $\dgamMC_{\geq
  4}$.

We choose starting scales of $\Tau_2^\cut$ and $\Tau_3^\cut$ for the
$\Phi_2$ and $\Phi_3$ events respectively. For the $\Phi_4$ events, the starting
scale $t$ needs to be  a measure of the hardness of the splitting, for example the \threej value $\Tau_3$. Here we follow the choice made in Ref.~\cite{Bizon:2019tfo} and set 
\begin{equation}
  t=2p_{\mathrm{daughter}}\cdot p_{\mathrm{sister}}\frac{E_{\mathrm{daughter}}}{E_{\mathrm{sister}}}\,,
\end{equation}
where the energies are defined in the Higgs boson rest frame.
After interfacing to the \pythiaEight parton shower, we expect the accuracy
for observables other than $\mathcal{T}_2$ to be no worse than that of the standalone \pythiaEight shower.

\subsubsection{Nonperturbative power corrections and hadronisation}

The approach described up to now does not take into account
nonperturbative power corrections, which can significantly affect the
partonic predictions.  The framework of SCET allows these
nonperturbative effects to be systematically included via the
introduction of a shape function $f(k, \mu)$ modifying the soft
component as \cite{Korchemsky:1999kt,Hoang:2007vb,Ligeti:2008ac}
\begin{equation}
S_2(\Tau_2, \mu) = \int \df k \, S_2^{\rm pert}(\Tau_2 - k, \mu) \, f(k, \mu),
\end{equation}
where $ S_2^{\rm pert}$ is the perturbative soft function.
At small $\Tau_2 \sim  \lqcd$ the shape function
gives an $\ord{1}$ contribution to the cross section, while for larger $\Tau_2$ values 
one can show that the dominant contribution is of $\ord{\lqcd/\Tau_2}$ and results in a overall shift of
 the $\Tau_2$ spectrum~\cite{Abbate:2010xh}. The same conclusions can be reached using a dispersive model
and an effective value for the strong coupling constant in the nonperturbative regime~\cite{Dokshitzer:1995zt,Dokshitzer:1995qm, Dokshitzer:1998kz, Gehrmann:2012sc}.

The resummed
predictions obtained by \geneva  at the partonic level only include the perturbative soft function,
 and we delegate the provision of nonperturbative ingredients to the hadronisation models used in \pythiaEight.
Therefore, after the showering stage, the events are interfaced to the phenomenological hadronisation model in \pythiaEight
 without further constraints on the kinematics of the hadronised event. This means that the hadronisation can potentially cause significant shifts of the $\Tau_2$ spectrum.

It is known that the \twoj and the thrust observables receive
different hadronisation corrections, due to the different treatment of the
hadron masses in their definitions~\cite{Salam:2001bd,Mateu:2012nk}.
Since there are currently no experimental data with which we can compare for these decay channels, in this work we consistently use the definition of $\Tau_2$ in \eq{Tau2def} even for hadronised events, despite the larger power corrections compared to schemes with a different mass treatment. This is different from the approach taken for the $e^+ e^- \to $ jets study in Ref.~\cite{Alioli:2012fc}, where the definition based on thrust was used to compare to LEP data.

It is important to notice that we do not include uncertainties from
these nonperturbative contributions in the results presented in the
next section.  In our approach, a crude estimate of their size could
in principle be obtained by varying the tune parameters of the
\pythiaEight hadronisation model, but a more detailed study of this
goes beyond the scope of this work.
It is worth noting, however, that in any calculation obtained by matching higher-order calculations with parton shower one has to carefully evaluate which parameters are truly encoding nonperturbative effects and should therefore be tuned.

\section{Results}
\label{sec:genevaresults}
In this section we present the full \geneva results obtained by
matching the resummed calculation to the fixed order. We adopt the
same values of SM parameters as in \subsec{resumresults} and set
$\Tau_2^\cut=\Tau_3^\cut=1\GeV$. We interface to the \pythiaEight
generator which showers our events\footnote{The publicly available
  \pythiaEight.235 version we used has difficulty parsing events read
  from an LHEF file in which only one particle appears in the initial
  state -- we therefore add dummy neutrino beams using code provided
  by S.~Prestel to mimic a collider process.} and use the $e^+e^-$
tune 3, turn off QED effects and prevent the decay of $b$-hadrons. We
set the strong coupling used by \pythiaEight to $\as=0.118$, although ideally, one should perform a dedicated tune to accommodate for this change.

With the setup as described, we verified that we obtain the correct
NNLO decay rate up to the power corrections shown in \fig{nscum}. The
partonic results are presented for each channel in \tab{decrates}, where the
analytic values have been obtained using the formulae appearing in
\app{NNLOdecrates}.
\begin{table}[t]
\centering
\begin{tabular}{ |c|c|c| }
\hline
$\Gamma^i_{\mathrm{NNLO}}(\mu)/\GeV$ & Analytic & \geneva, $\Tau_2^\cut=1\GeV$ \\
\hhline{|=|=|=|}
$i=b,\,\mu=M_H$ & \num{3.053e-3} & \num{3.042e-3}$\pm$\num{0.2e-5} \\
\hline
$i=b,\,\mu=2M_H$ & \num{3.104e-3} & \num{3.094e-3}$\pm$\num{0.2e-5} \\
\hline
$i=b,\,\mu=M_H/2$ & \num{2.973e-3} & \num{2.961e-3}$\pm$\num{0.2e-5} \\ 
\hline
$i=g,\,\mu=M_H$ & \num{3.374e-4} & \num{3.338e-4}$\pm$\num{1.2e-6} \\
\hline
$i=g,\,\mu=2M_H$ & \num{3.189e-4} & \num{3.178e-4}$\pm$\num{0.8e-6} \\
\hline
$i=g,\,\mu=M_H/2$ & \num{3.491e-4} & \num{3.407e-4}$\pm$\num{1.8e-6} \\
\hline
\end{tabular}

\caption{ Comparison of Higgs boson partial widths
  obtained from NNLO analytic expressions and at the partonic level from \geneva. Note that,
  due to the presence of the power corrections displayed in
  \fig{nscum}, the values do not agree exactly within the statistical
  error and therefore a reweighting must be performed.  }
\label{tab:decrates}
\end{table}
In general, the \geneva method also guarantees
  NNLO accuracy for distributions differential in the Born variables
  of the process (see for example Ref.~\cite{Alioli:2019qzz}). In the
  case of a spin-0 boson decaying into two particles, however, the
  Born phase space is parameterised by only two angles and is flat in
  both -- there is therefore no non-trivial shape information which
  can be compared to a fixed-order calculation. We have, however,
  validated our NLO calculations of $H\rightarrow b\bar{b}g$ and
  $H\rightarrow ggg$/$H\rightarrow q\bar{q}g$ against
  \amcatnlo~\cite{Alwall:2014hca} and found perfect agreement. We
  checked that by increasing the $\Tau_2^\cut$ to $\sim5$ GeV we
  obtain smaller power corrections (see \fig{nscum}) -- however, since
  this would limit our higher-order resummed predictions for the shape
  of the spectrum to $\Tau_2> 5 \GeV$, in the following we continue to
  use $\Tau_2^\cut=1\GeV$ and accordingly reweight our events in order
  to obtain the correct total decay width.

\begin{figure*}[t]
  \begin{subfigure}[b]{\rescalethreeplots}
    \includegraphics[width=\textwidth]{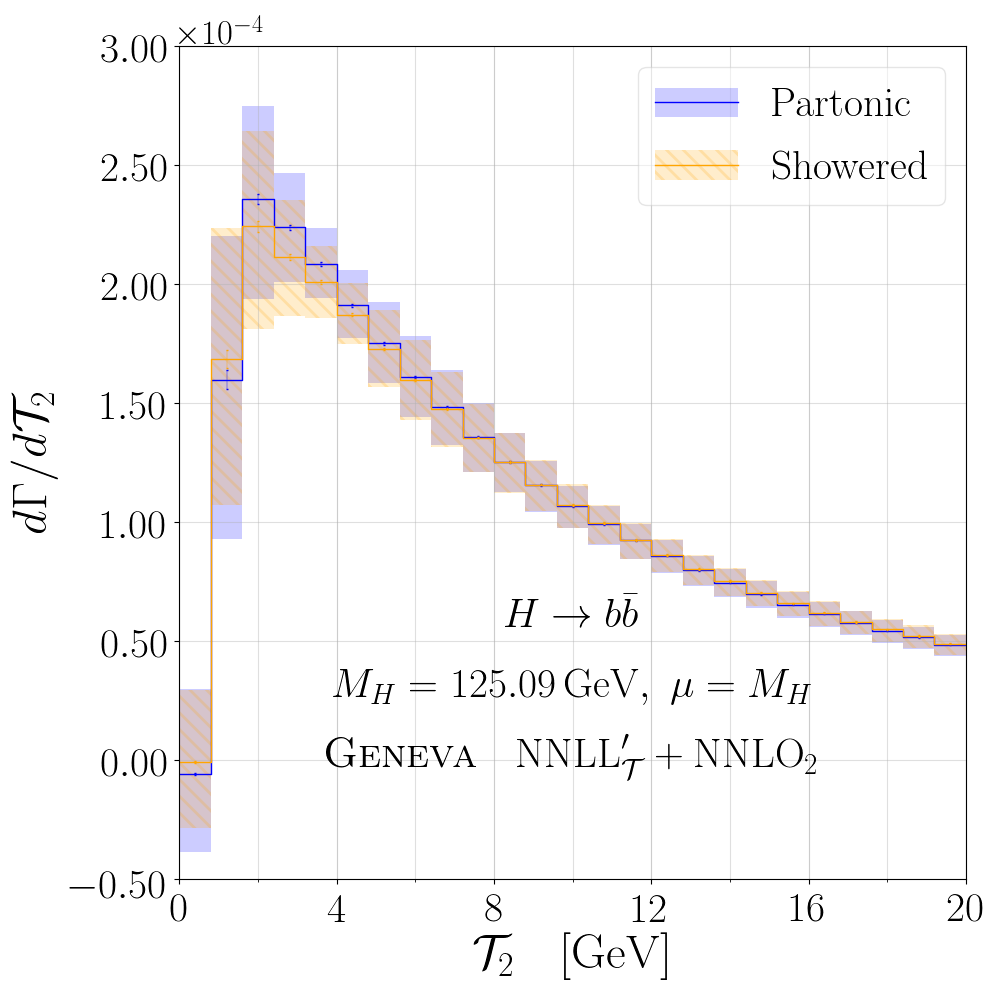}%
    \label{fig:Tau2peakbb}
  \end{subfigure}
  \begin{subfigure}[b]{\rescalethreeplots}
    \includegraphics[width=\textwidth]{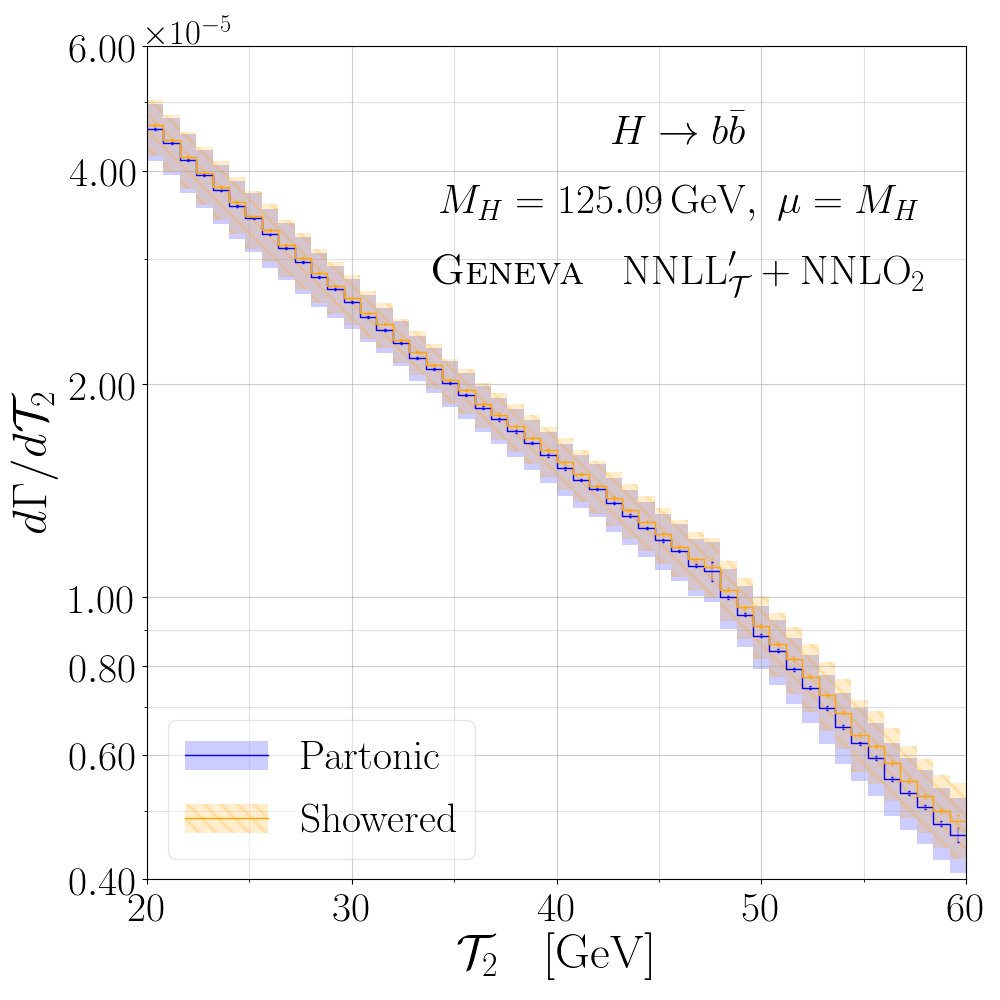}%
    \label{fig:Tau2transbb}
  \end{subfigure}
  \begin{subfigure}[b]{\rescalethreeplots}
    \includegraphics[width=\textwidth]{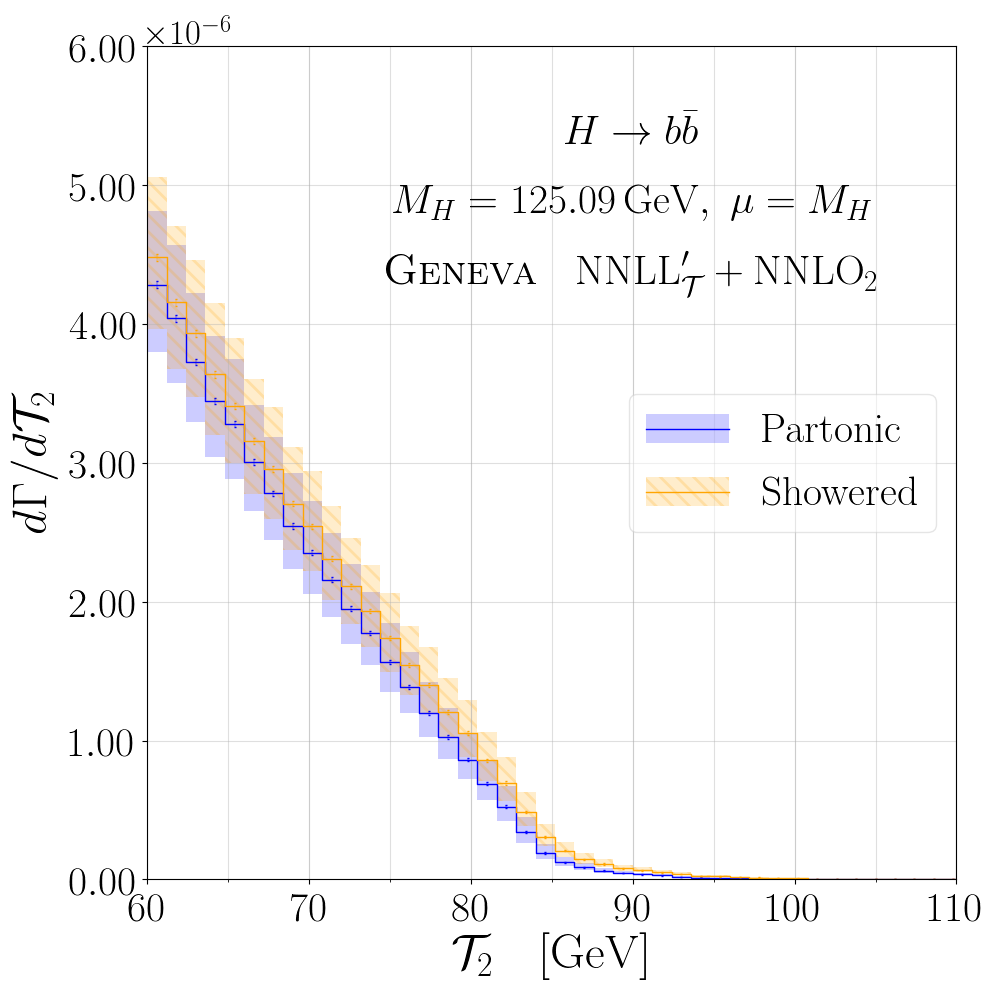}%
    \label{fig:Tau2tailbb}
  \end{subfigure}
  \vspace{\spacebeforefigurecaption}
  \caption{Validation of the $\Tau_2$ spectrum in \geneva for
    $H\rightarrow b\bar{b}$. The partonic NNLL$'$+NNLO $\Tau_2$
    resummation is compared to the showered results, before the
    addition of nonperturbative effects.}
\label{fig:tau2validationbb}
\vspace*{4ex}
  \begin{subfigure}[b]{\rescalethreeplots}
    \includegraphics[width=\textwidth]{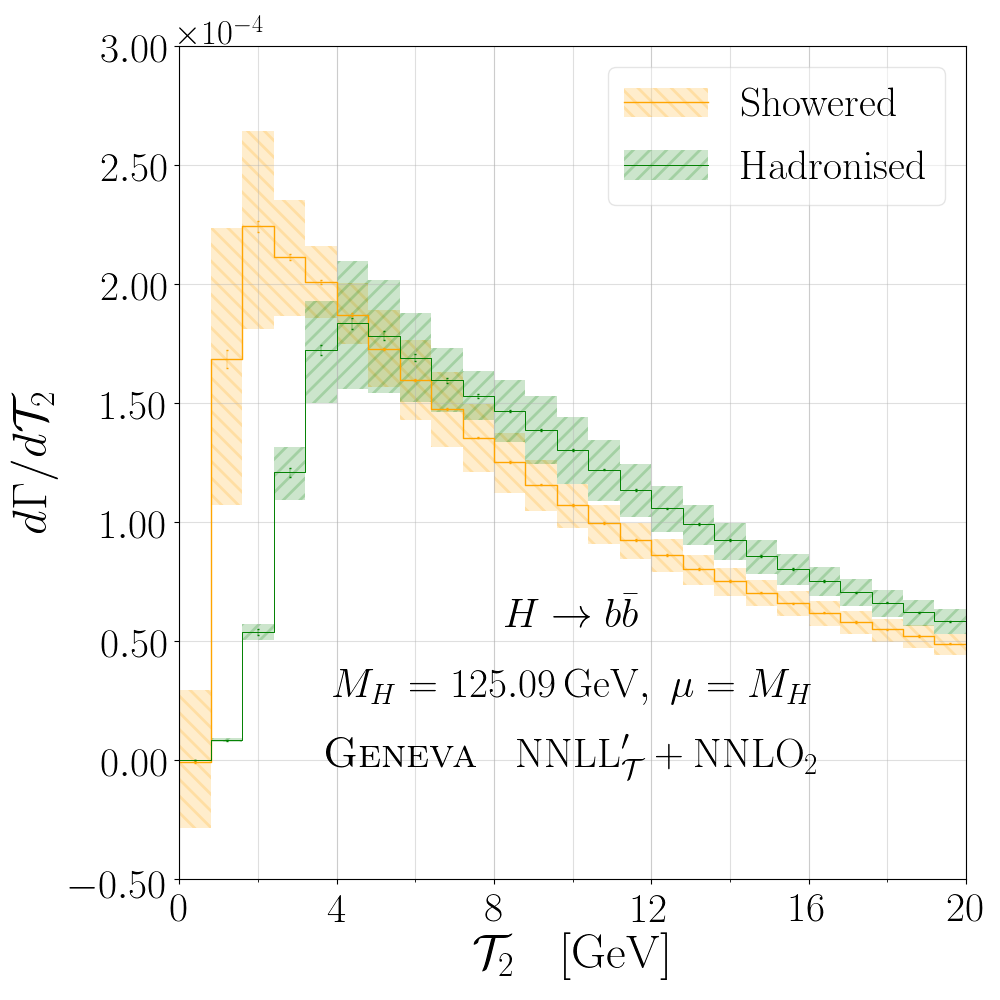}%
    \label{fig:Tau2peakhadbb}
  \end{subfigure}
  \hspace*{\hspacebetweenthreeplots}
  \begin{subfigure}[b]{\rescalethreeplots}
    \includegraphics[width=\textwidth]{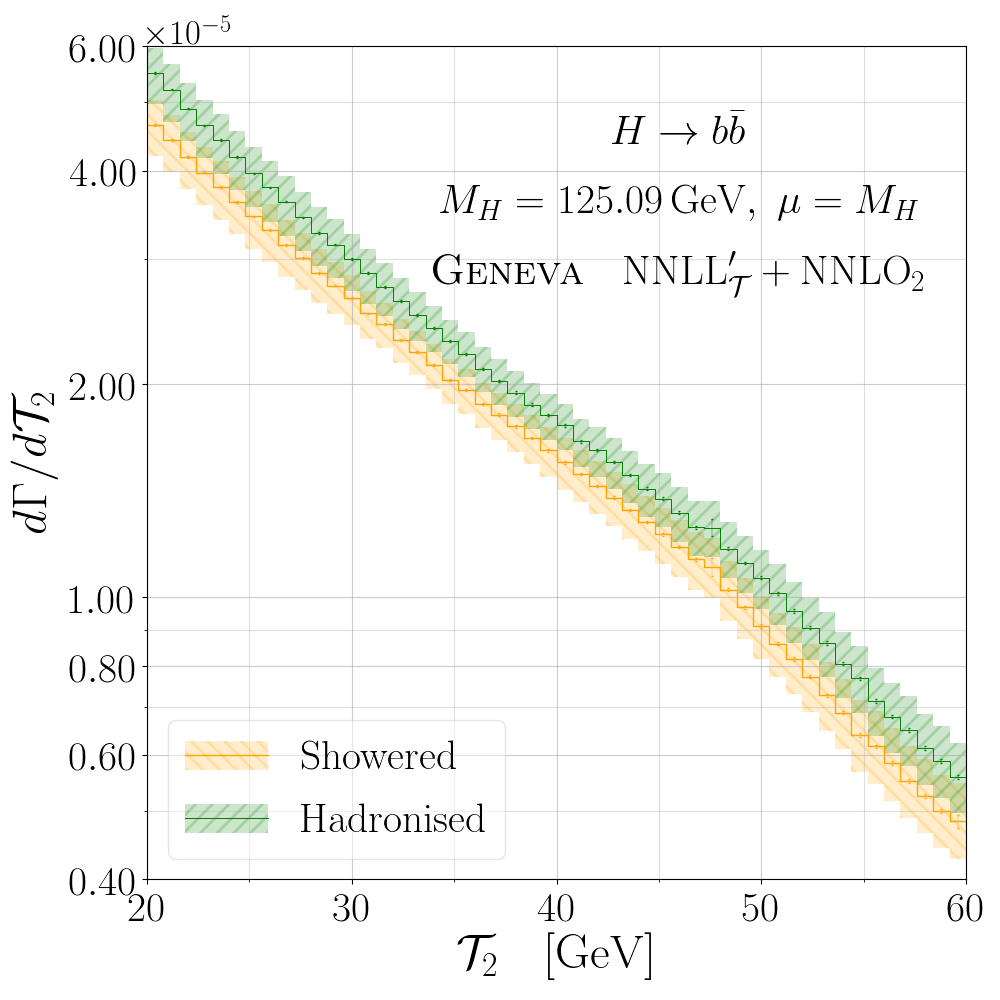}%
    \label{fig:Tau2transhadbb}
  \end{subfigure}
  \hspace*{\hspacebetweenthreeplots}
  \begin{subfigure}[b]{\rescalethreeplots}
    \includegraphics[width=\textwidth]{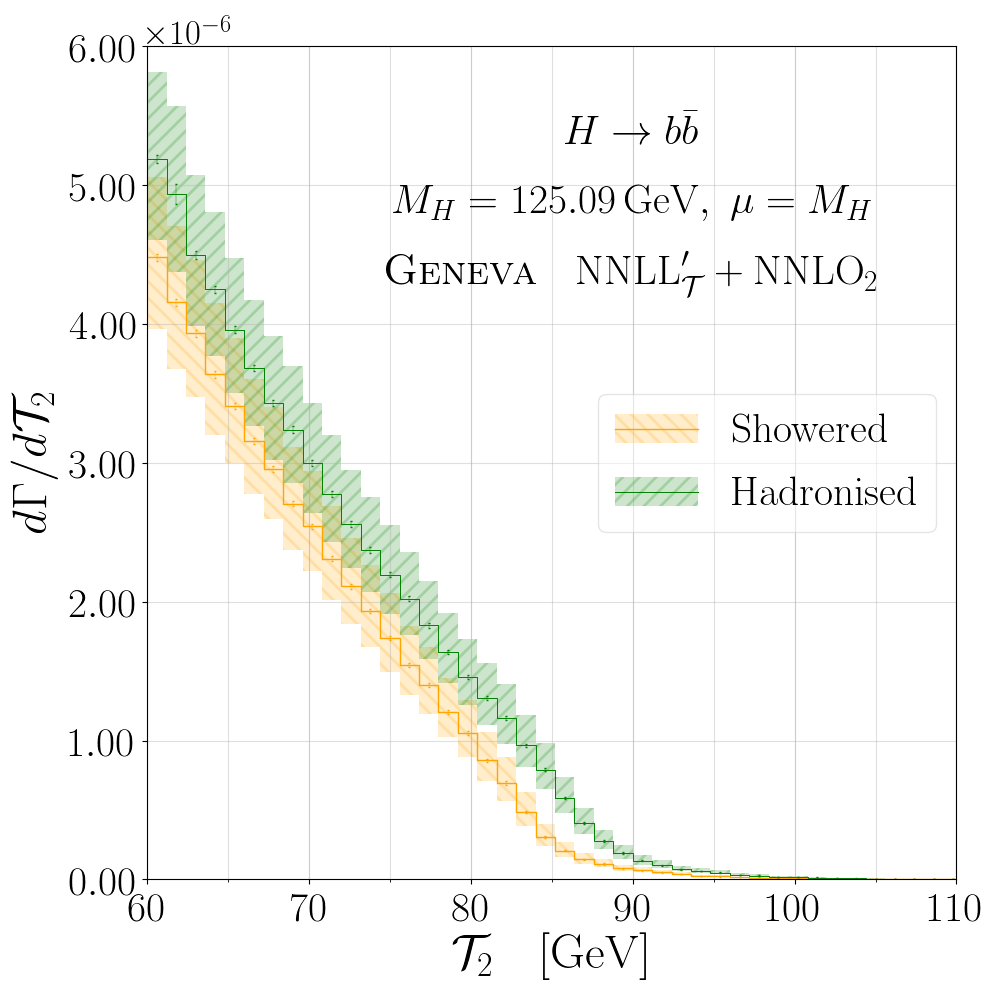}%
    \label{fig:Tau2tailhadbb}
  \end{subfigure}
  \vspace{\spacebeforefigurecaption}
  \caption{Comparison of the showered and hadronised $\Tau_2$ spectra in \geneva for $H\rightarrow b\bar{b}$.}
\label{fig:tau2hadvalidationbb}
\end{figure*}

\begin{figure*}[t]
  \begin{subfigure}[b]{\rescalethreeplots}
    \includegraphics[width=\textwidth]{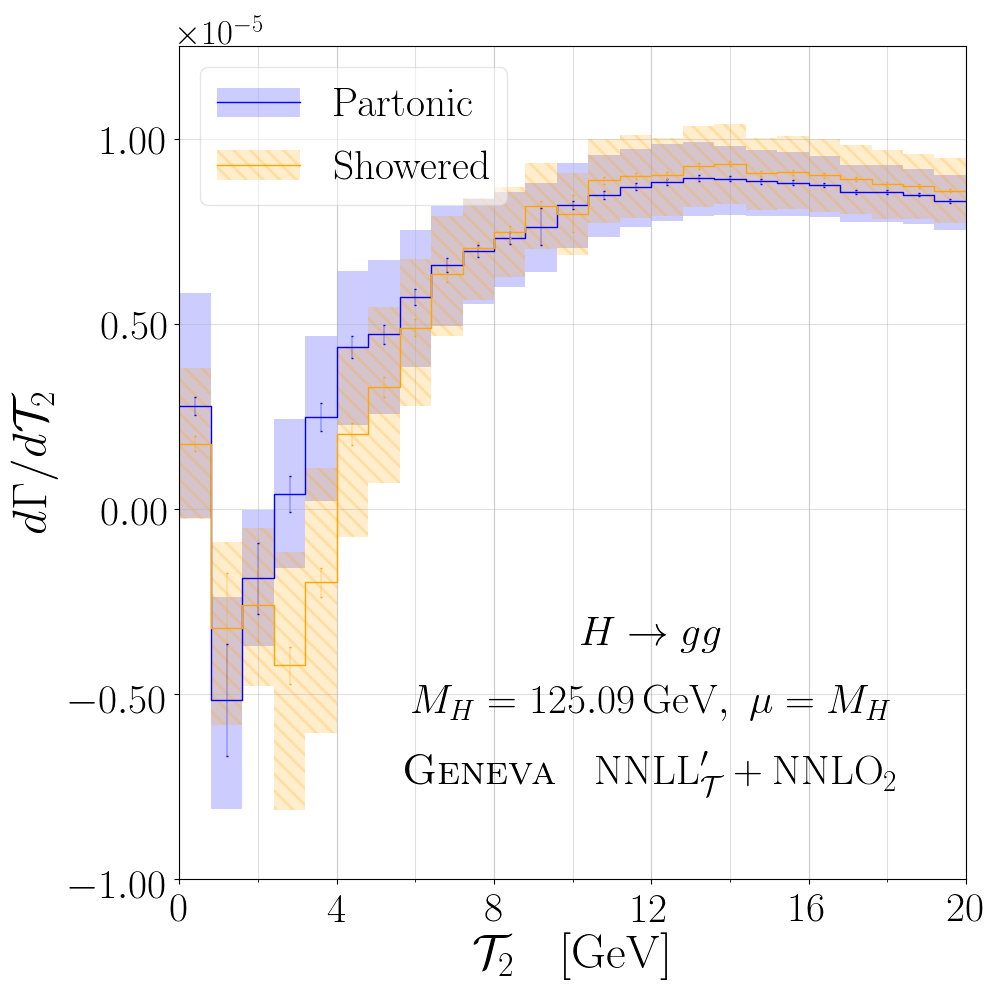}%
    \label{fig:Tau2peakgg}
  \end{subfigure}
  \hspace*{\hspacebetweenthreeplots}
  \begin{subfigure}[b]{\rescalethreeplots}
    \includegraphics[width=\textwidth]{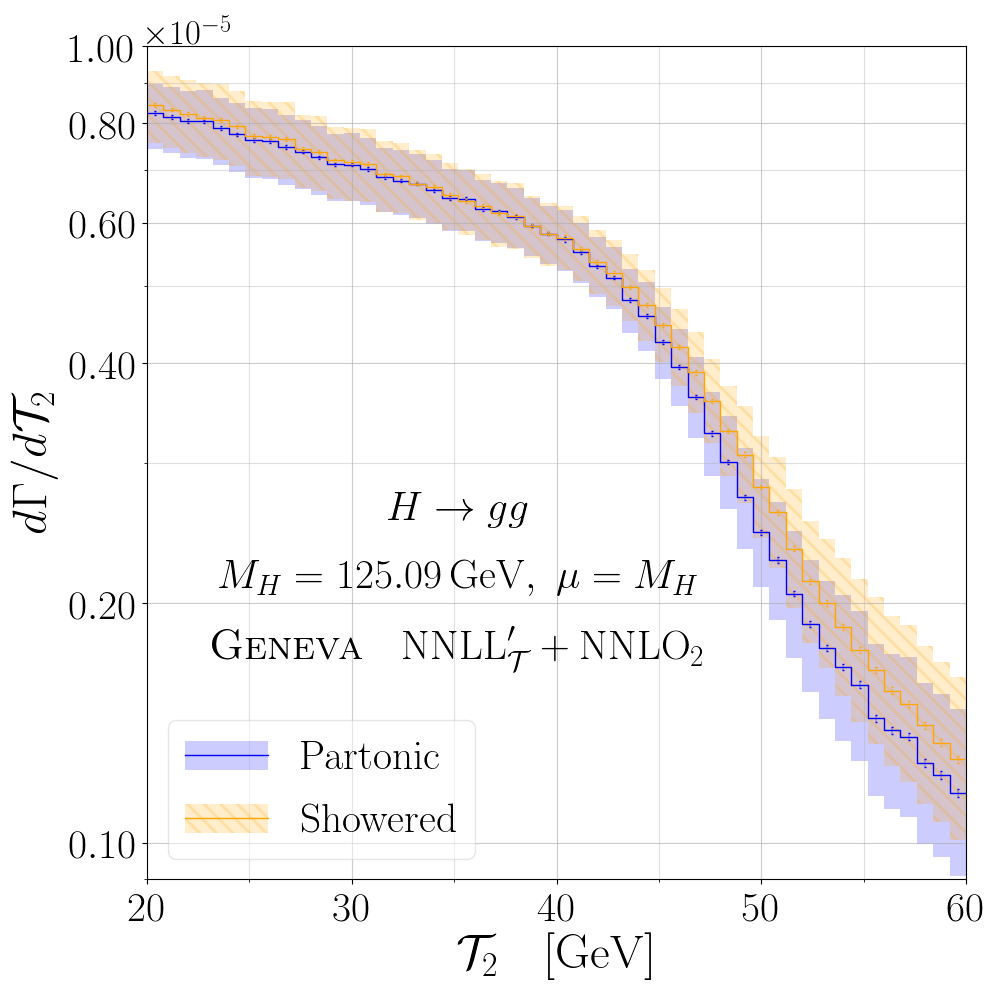}%
    \label{fig:Tau2transgg}
  \end{subfigure}
  \hspace*{\hspacebetweenthreeplots}
  \begin{subfigure}[b]{\rescalethreeplots}
    \includegraphics[width=\textwidth]{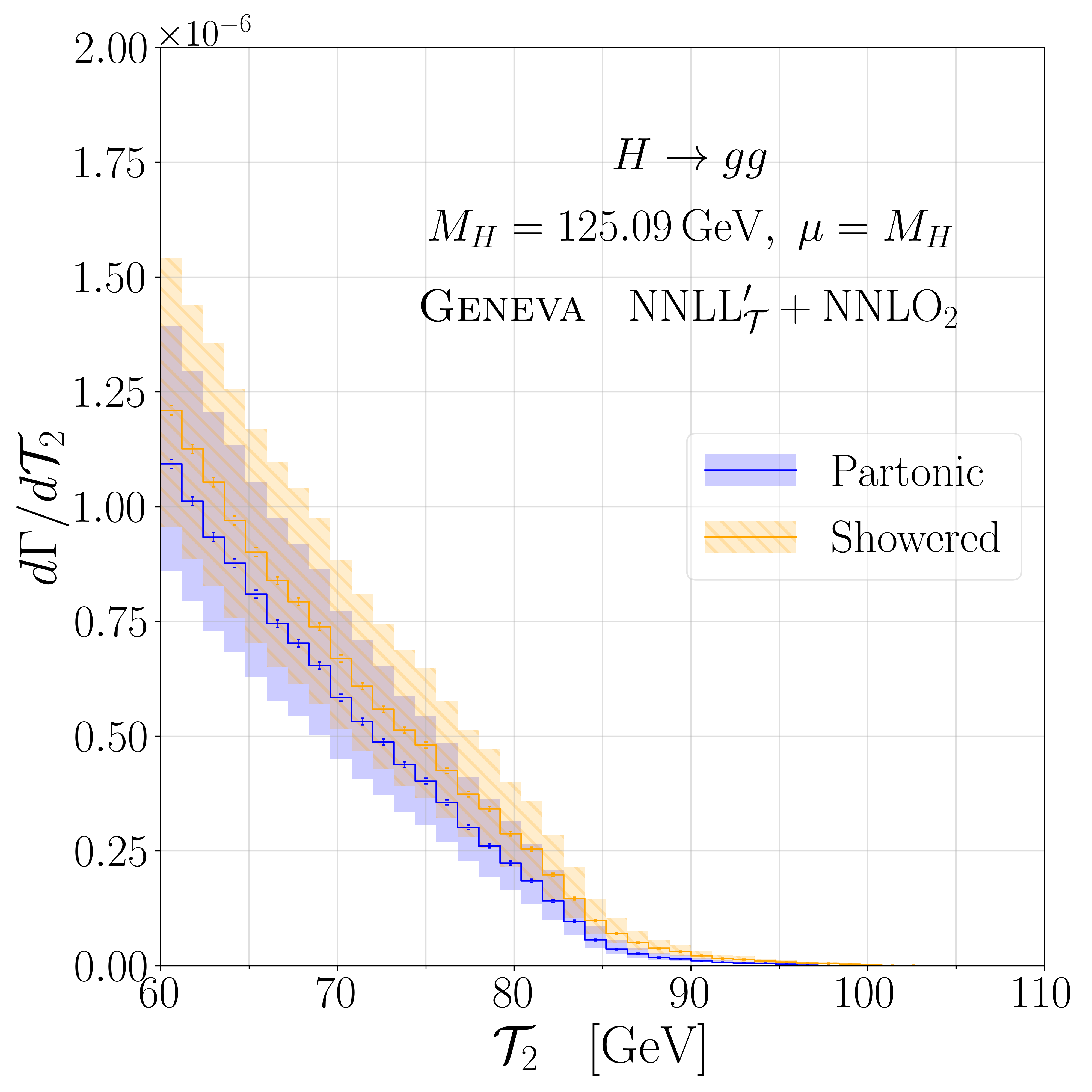}%
    \label{fig:Tau2tailgg}
  \end{subfigure}
  \vspace{\spacebeforefigurecaption}
  \caption{Validation of the $\Tau_2$ spectrum in \geneva for
    $H\rightarrow gg$. The partonic NNLL$'$+NNLO $\Tau_2$ resummation
    is compared to the showered results, before the addition of
    nonperturbative effects.}
\label{fig:tau2validationgg}
\vspace*{4ex}
  \begin{subfigure}[b]{\rescalethreeplots}
    \includegraphics[width=\textwidth]{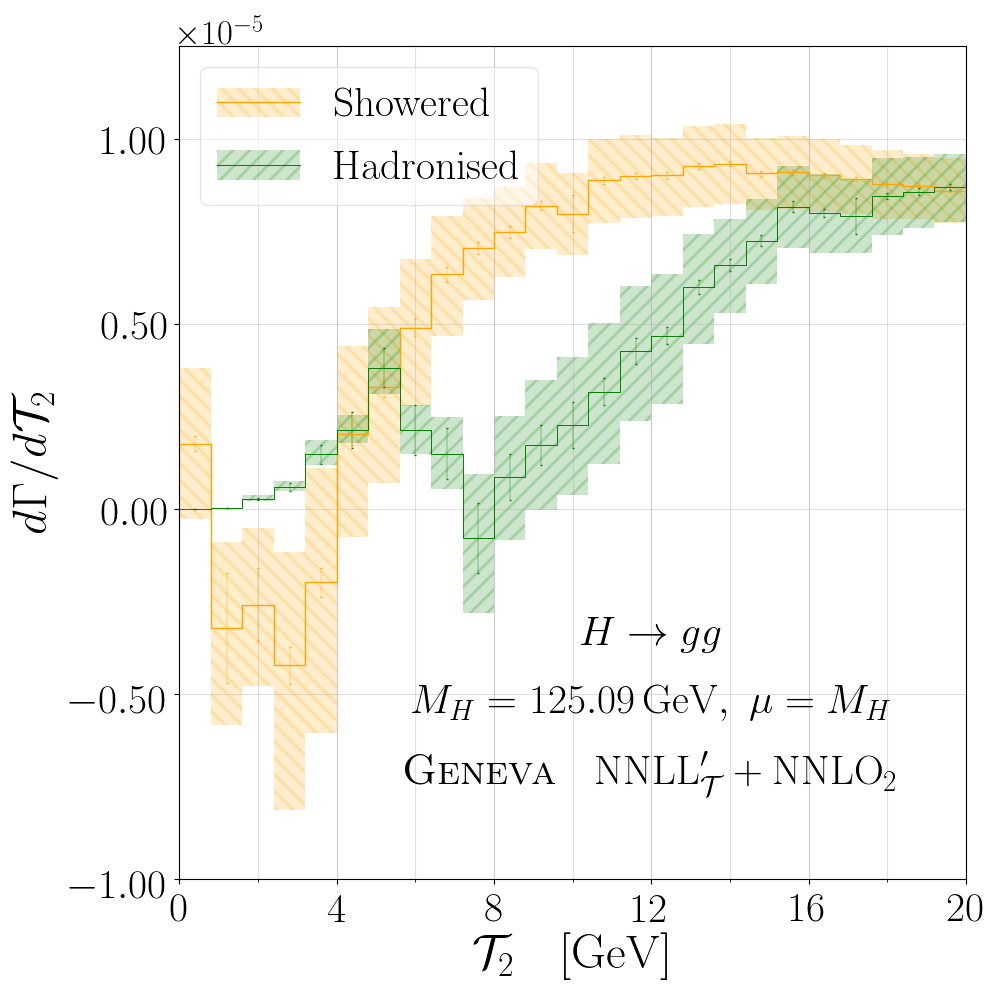}%
    \label{fig:Tau2peakhadgg}
  \end{subfigure}
  \hspace*{\hspacebetweenthreeplots}
  \begin{subfigure}[b]{\rescalethreeplots}
    \includegraphics[width=\textwidth]{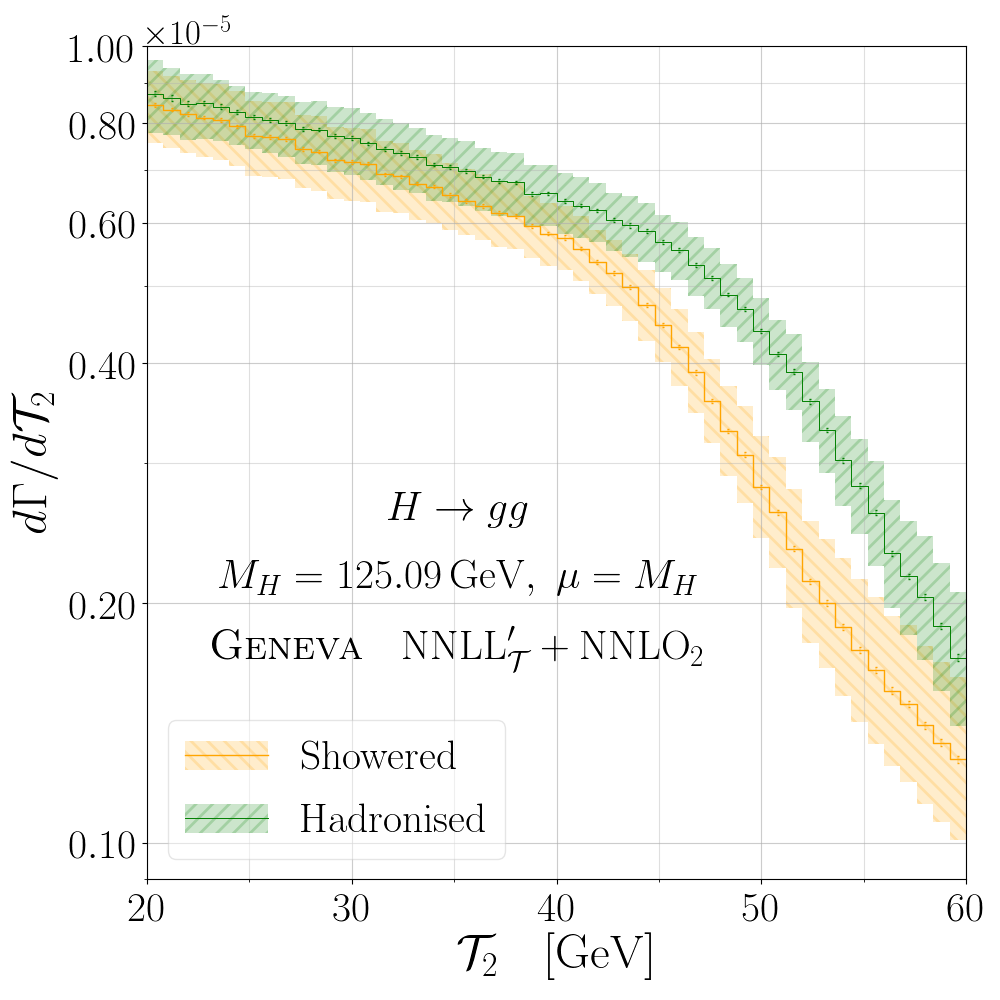}%
    \label{fig:Tau2transhadgg}
  \end{subfigure}
  \hspace*{\hspacebetweenthreeplots}
  \begin{subfigure}[b]{\rescalethreeplots}
    \includegraphics[width=\textwidth]{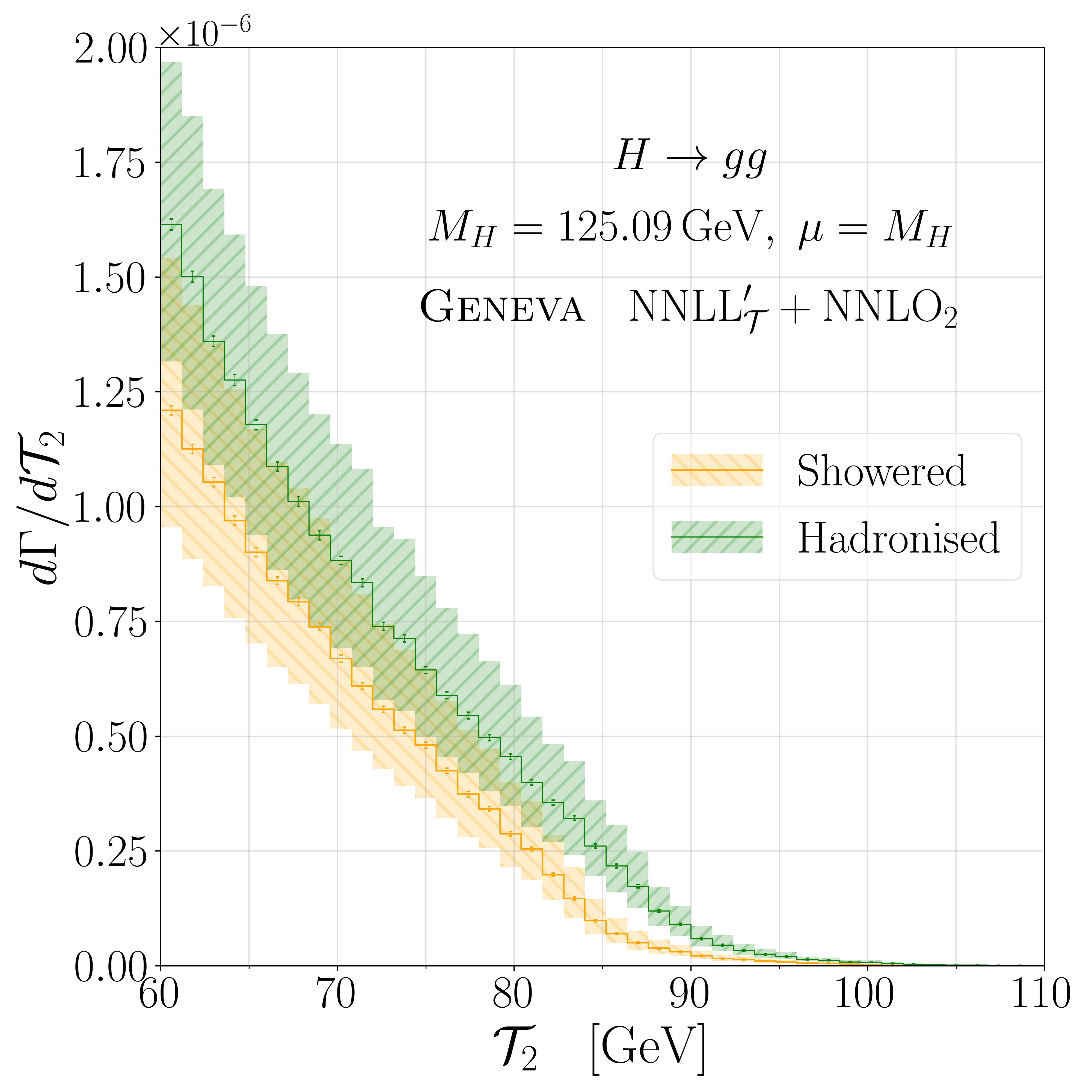}%
    \label{fig:Tau2tailhadgg}
  \end{subfigure}
  \vspace{\spacebeforefigurecaption}
  \caption{Comparison of the showered and hadronised $\Tau_2$ spectra in \geneva for $H\rightarrow gg$.}
\label{fig:tau2hadvalidationgg}
\end{figure*}

A comparison of the NNLO+NNLL$'$ results at the partonic and showered
levels is presented in \fig{tau2validationbb} for the
$H\rightarrow b\bar{b}$ process and in \fig{tau2validationgg} for the
$H\rightarrow gg$ process, while the corresponding comparisons of the
showered and hadronised events are shown in
\figs{tau2hadvalidationbb}{tau2hadvalidationgg}. The panels in the
plots show three different regions of the \twoj spectrum: the peak
(leftmost panels), where resummation effects are expected to be
dominant; the transition (centre panels), where the resummed and fixed-order
calculations compete for importance; and the tail (rightmost
panels), where the resummation is switched off and the fixed-order
calculation provides the correct physical description. We observe that
in the $b\bar{b}$ channel the $\Tau_2$ is well preserved by the
shower, while hadronisation effects shift the distribution to higher
values of $\Tau_2$ across all regions. This can be compared to the
results obtained in Ref.~\cite{Alioli:2012fc}, keeping in mind the
aforementioned difference between the \twoj definitions used at hadron
level and the different energy scale which result in competing contributions
to the shift.

In the $gg$ channel, the effect of the shower on the $\Tau_2$ spectrum
is greater, especially at the lowest values of $\Tau_2$, but it
preserves the shape of the distribution to within the scale variation
bands closer to the peak and in the transition region. We also notice
that the partonic and showered predictions give a negative cross
section for very small values of $\Tau_2$, below the nonperturbative
freeze-out of the profile scales. This behaviour should not be
concerning as it happens in a region where the perturbative resummed
results are already questionable  and, as mentioned before, we do not
include any nonperturbative uncertainty. We have verified that the
size of the negative value is augmented by the nonsingular corrections
which we include via an additive approach. Indeed, when examining
the resummed $\Tau_2$ distribution alone, the behaviour at small
$\Tau_2$ remains negative but is compatible with a value of zero
to within the quoted uncertainties.

A peculiar feature is observed in the first bin of \fig{tau2validationgg}, which contains the
cross section below $\Tau_2^{\rm cut}$ and is positive. This is again a
consequence of the missing nonsingular corrections in \eq{2tilde},
which are included by the reweighting procedure.  Since these are
particularly large for this process, see \fig{nscum}, their effect is
to change the sign of the cumulant below $\Tau_2^{\rm cut}$.

We also observe somewhat larger effects on the spectrum due to
hadronisation compared to the $b\bar{b}$ case, particularly in the
peak region. The seemingly unusual behaviour at small $\Tau_2$ is a
consequence of the already discussed shift of the spectrum after
hadronisation resulting in a smearing of the first bin.  We stress
again that the small error bands reported are due to the lack of
nonperturbative uncertainties.

\begin{figure}
  \centering
  \includegraphics[width=0.48\textwidth]{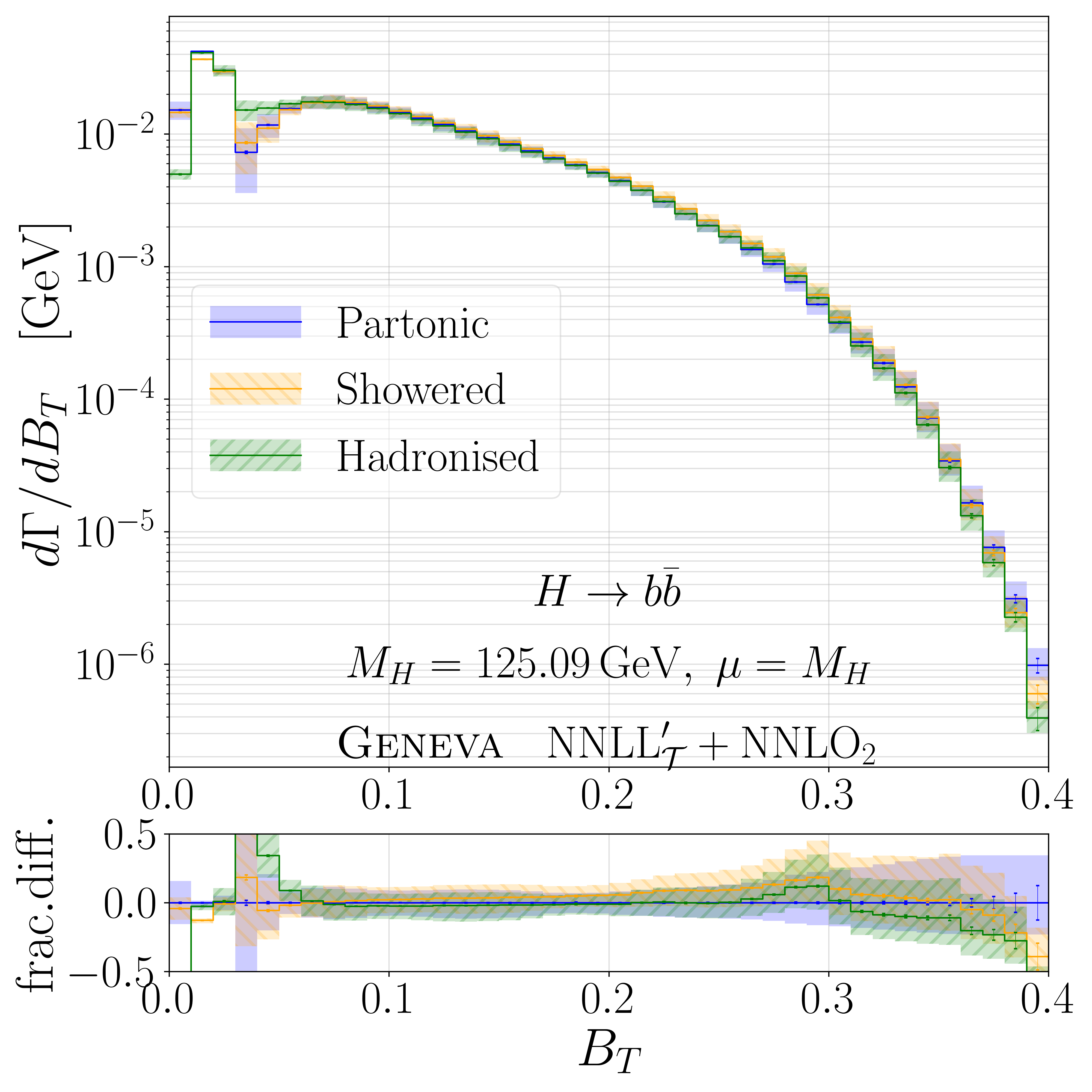}
  \includegraphics[width=0.48\textwidth]{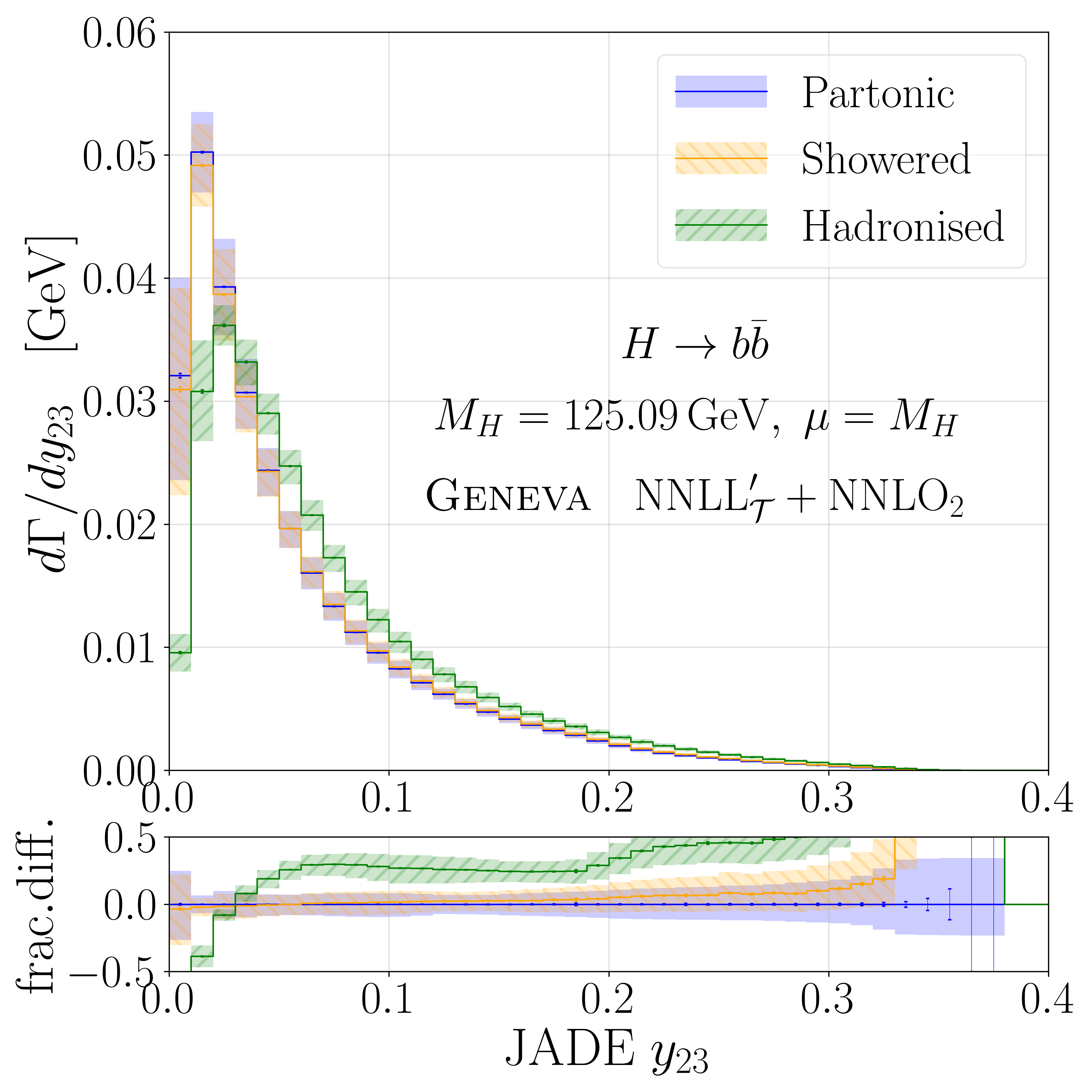}
  \caption{Jet broadening and the JADE two-to-three differential jet rate at the partonic, showered and hadronised levels for $H\to b\bar{b}$.}
  \label{fig:shapesbb}
\vspace*{4ex}
  \centering
  \includegraphics[width=0.48\textwidth]{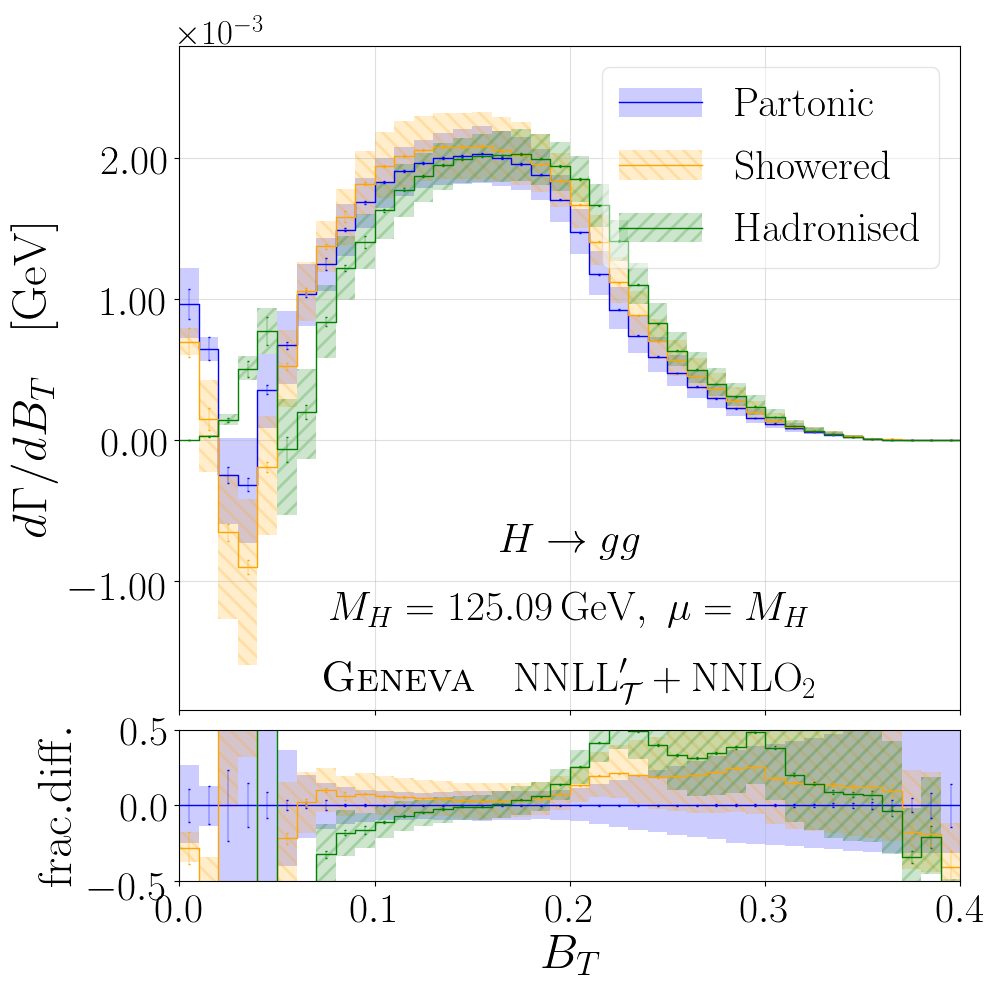}
  \includegraphics[width=0.48\textwidth]{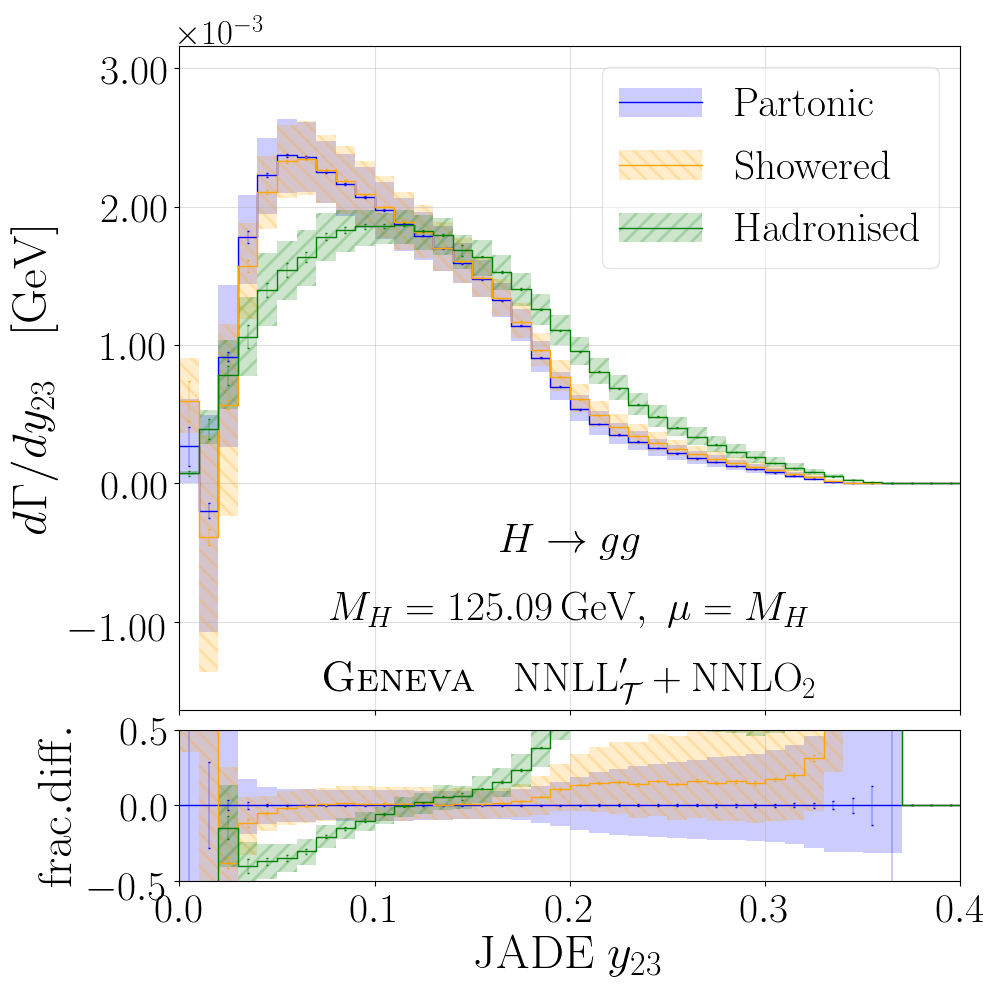}
  \caption{Jet broadening and the JADE two-to-three differential jet rate at the partonic, showered and hadronised levels for $H\to gg$.}
  \label{fig:shapesgg}
\end{figure}
Finally, in \figs{shapesbb}{shapesgg} we show the results for
distributions other than the \twoj that we use as input to our \geneva
implementation, for the $b\bar b$ and $gg$ cases respectively.  We
consider the JADE clustering metric $y_{23}$ for separating two exclusive
jets from three or more~\cite{Bartel:1982ub,Bartel:1986ua} and the jet broadening ($B_T$)~\cite{Rakow:1981qn,Catani:1992jc}
event shape defined as follows
\begin{align}\label{eq:defbt}
B_T &= \frac{1}{2\sum_k \abs{\vec{p}_k } } \sum_i \abs{ \vec{p}_i \times \hat{n}_T } \,,
\end{align}
where the sum runs over all final state particles and $\hat{n}_T$ is the thrust axis.

It is important to remark that we do not expect the \geneva method to
provide a higher formal accuracy for these observables, but it is
nonetheless interesting to observe the effects of our predictions at
the various stages.  In general the $b\bar b$ decay channel is better
behaved after showering, providing results that are compatible with
the predictions at the partonic level over the majority of the phase
space.

We notice that, at small values of the jet broadening, some unappealing artefacts appear. We have verified that these features are a consequence of the additive matching to the fixed-order calculation, used to include the nonsingular corrections. They are not present, for example, when examining the jet broadening distribution of events obtained from the $\Tau_2$ resummed calculation alone. We therefore conclude that the secondary peak which appears constitutes an effect beyond the perturbative accuracy of our formulae, despite being numerically large. In the future, one could explore whether by exploiting the freedom in the handling of the recoil by the $3\to4$ mapping one could ameliorate this undesirable effect -- however, we do not pursue this further here. We also see deviations for particular values of the
observables after hadronisation and hadron decays are included. In
particular we notice a significant shift in the JADE $y_{23}$
observable for both decay channels, which is not unexpected for this
specific jet-clustering measure.
\section{Conclusions}
\label{sec:conc}
In this work we have resummed the \twoj at NNLL$'$ for hadronic Higgs
boson decays in the $b\bar{b}$ and $gg$ channels via a SCET
approach. Compared to previous fixed-order results, we observe the
expected improved behaviour in the small $\Tau_2$ region, where the
physical Sudakov peak is now described correctly.  We have also
implemented these processes in the \geneva framework, which has
allowed us to match the resummed calculations with NNLO fixed-order
predictions and to a parton shower. This has required an examination
of the interplay of the singular and nonsingular contributions, in
order to determine the region in which resummation effects are
dominant and hence design profile scales which provide a smooth
transition between the resummed and fixed-order regimes. As a result
we have produced NNLO accurate event generators interfaced to the
\pythiaEight parton shower for the two processes, which provide
accurate predictions in all regions of phase space.

We compared predictions at the partonic, showered and hadronised
levels, finding the expected good agreement for the total decay rates
and for the $\Tau_2$  distribution up to the showered level.
We observed larger differences due to the hadronisation, especially in the
$gg$ channel.

The completion of this work will eventually allow us to combine our
$H\rightarrow b\bar{b}$ result with the \geneva $VH$ production
generator in the narrow width approximation, yielding a full NNLOPS
generator for the signal channel of the $l^+l^{-}b\bar{b}$ final
state. Given the recent
observation of the Higgsstrahlung process by the ATLAS \& CMS experiments at the LHC~\cite{Sirunyan:2018kst,Aaboud:2018zhk},
this will constitute an important phenomenological result. It will
also allow a direct comparison with the only other existing NNLOPS
generator for this process~\cite{Bizon:2019tfo}. In light of the
findings in Ref.~\cite{Gao:2019mlt} regarding the convergence of the
perturbative series and the N\textsuperscript{3}LO results at fixed
order which are also available for this decay channel, it might also
be interesting to consider building an event generator at
N\textsuperscript{3}LOPS level. Another avenue for development might
be the inclusion of a finite $b$-quark mass in the
generator, given recent work on fixed-order
calculations~\cite{Behring:2020uzq}. We leave this to future
consideration.

\acknowledgments We are grateful to Marco Zaro for help in comparing
to \amcatnlo fixed-order results and to Stefan Prestel for interfacing
the LHEF files with Pythia8. MAL thanks HexagonFab, Cambridge for
hospitality during the completion of this work. The work of SA, AB,
SK, RN, DN and LR is supported by the ERC Starting Grant
REINVENT-714788.  SA and MAL acknowledge funding from Fondazione
Cariplo and Regione Lombardia, grant 2017-2070. The work of SA is also
supported by MIUR through the FARE grant R18ZRBEAFC. We acknowledge
the CINECA award under the ISCRA initiative and the  National Energy Research Scientific Computing Center (NERSC), a U.S. Department of Energy Office of
Science User Facility operated under Contract No. DEAC02-05CH11231,
for the availability of
the high performance computing resources needed for this work.

\appendix
\section{Constructing a \twoj-preserving map}
\label{app:tau2map}
The map used for $3 \to 4$-body splittings and $4 \to 3$-body
projections presented in this section was first developed and applied
to the process $e^+e^-\rightarrow jj$ in
Ref.~\cite{Alioli:2012fc}. Here, we detail the construction of the map
as used in that work and in addition provide the translation to the
splitting variables $\Tau_2,\,z$ and $\phi$ needed for the Higgs boson
decay case.

We start by considering the case of a splitting, which takes as input
$N$-body phase space points $\Phi_N$ and generates from them
$(N\!+\!1)$-body phase-space points $\Phi_{N+1}$. Since we wish to
calculate the NLO distribution in \twoj, $\Tau_2$, while still
generating exclusive $\Phi_3$ points, we must use a map that produces
$\Phi_4$ points with the same value of $\Tau_2$ as the $\Phi_3$ points
with which we started. Unfortunately, the construction of such a map
is challenging since $\Tau_2$ is a global variable. A more manageable
approach is to seek a map which preserves not the exact \twoj,
$\Tau_2$, but instead a related quantity, the fully recursive \twoj,
$\Tau_2^{\mathrm{FR}}$, defined by the following procedure:
\begin{enumerate}
\item Recursively cluster the starting phase space point $\Phi_M$ down to a $\Phi_3$ point using the \nj metric for final state particles
  \begin{equation}
    \rho_{ij}=|\vec{p}_i|+|\vec{p}_j|-|\vec{p}_i+\vec{p}_j|.
  \end{equation}
\item Measure $\Tau_2$ on the resulting $\Phi_3$ point.
\end{enumerate}
The quantity we obtain through this procedure has the same singular
structure as the exact $\Tau_2$, with any differences being captured
by the nonsingular contributions.

Starting from the $3$-parton phase space point $\Phi_3$, which is the
input of the splitting map, we label its momenta as
\begin{equation}
  p_1,\,p_2,\,p_3\quad \mathrm{with}\quad p_1^0>p_2^0>p_3^0\,.
\end{equation}
The thrust axis will lie along the direction of the hardest parton
(i.e. along $\vec{p}_1$), and we have that
\begin{equation}
  \label{eq:taubefore}
  \Taufr=\Tau_2=2\Ecm-4|\vec{p}_1|\,.
\end{equation}
When we split to a $4$-parton event and then cluster back to a massive
$3$-parton event, the thrust axis is still determined by the most
energetic of the three partons and we have that
\begin{equation}
  \label{eq:tauafter}
  \Taufr=2\Ecm-4|\vec{p}_{\mathrm{max}}|\,.
\end{equation}
If we are to preserve $\Taufr$, clearly \eqs{taubefore}{tauafter} must
be equal and so the hardest parton in the massive $3$-parton event
obtained after reclustering must be parton 1. We can then split the
massive leg to produce a $4$-parton point. The emitter may or may not be
the hardest parton -- these two cases must be treated separately.

We will now detail how the $\Phi_{4}$ point is obtained from the
$\Phi_3$ point in the two separate cases while preserving $\Taufr$. In
addition, we will show in each case that taking the singular limits of
the Jacobian of the transformation reproduces the limits of the FKS
Jacobian and that our fixed-order subtractions in \eq{FOFKS} therefore
survive unaltered.
\subsection{Case 1: the emitter as hardest parton}
We deal first with the case in which the emitter is the hardest
parton, which we call the FR primary (FRp) map. In this case the
emitter is $p_1$ and, denoting the sum of the momenta of the split pair by $k$,
we must have that $\vec{k}\parallel\vec{p}_1$ in order to keep the
thrust axis in the same direction. We must also have
$|\vec{k}|=|\vec{p}_1|$. These conditions therefore fix the
sum of the three-momenta of the split pair.

In order to proceed with the actual construction of the split configuration we
use the same choice of variables as in the FKS approach, and therefore adopt a similar notation:
 we label
  the  momenta in $\Phi_3$ by $\bar{k}_1,\bar{k}_2,\bar{k}_3$, with  the emitter chosen as $\bar{k}_3$. For the
$\Phi_4$ momenta we use $k_1,\dots,k_4$ with the split pair 
$k_3,k_4$ and $k=k_3+k_4$. The recoil momenta are defined as
\begin{align}
\bar{k}_{\mathrm{rec}}&=\bar{k}_1+\bar{k}_2\,,& k_{\mathrm{rec}}&=k_1+k_2\,.
\end{align}

As discussed above, the splitting preserves the three-momentum of the
emitter which constrains the momentum of the split pair and the
recoil:
\begin{align}
\bar{k}_3&=(\bar{k}_3^0,\vec{\bar{k}}_3),& \bar{k}_{\mathrm{rec}}&=(\bar{k}^0_{\mathrm{rec}},-\vec{\bar{k}}_3)=(\Ecm-\bar{k}^0_3,-\vec{\bar{k}}_3)\,,\\
k&=(k^0,\vec{\bar{k}}_3),& k_{\mathrm{rec}}&=(k^0_{\mathrm{rec}},-\vec{\bar{k}}_3)=(\Ecm-k^0,-\vec{\bar{k}}_3)\,.
\end{align}
We must now determine $k^0$ and define the recoil constituents such
that they remain massless and sum to $k_{\mathrm{rec}}$. Since we have
that $|\vec{k}|=|\vec{\bar{k}}_3|=\bar{k}_3^0$, we may obtain an
expression for $k^0$:
\begin{equation}
k^2=(k^0)^2-|\vec{k}|^2=(k^0)^2-(\bar{k}^0_3)^2\,.
\end{equation}
Recalling the definitions of the FKS variables $\Phi_{\mathrm{rad}}^{\mathrm{FKS}} \equiv \{ \xi, y, \phi\}$
\begin{align}
  \label{eq:FKSvars}
  k_{4}^0&=\frac{1}{2}\Ecm\xi,& k^2=2k_3^0k_{4}^0(1-y)\,,
\end{align}
we may substitute in and solve the quadratic equation; one obtains
\begin{align}
  \label{eq:k0sol}
k^0&=\frac{1}{2}\Ecm\xi(1-y)\pm\left((\bar{k}^0_3)^2-\left(\frac{1}{2}\Ecm\xi\right)^2(1-y^2)\right)^{1/2}\,,\\
k^0_3&=-\frac{1}{2}\Ecm\xi y \pm\left((\bar{k}^0_3)^2-\left(\frac{1}{2}\Ecm\xi\right)^2(1-y^2)\right)^{1/2}\,.
\end{align}
Having determined $k^0$ in terms of $\xi$ and $y$, we must carefully
examine which solutions are kinematically allowed. The specific
 $\Phi_{\mathrm{rad}}^{\mathrm{FKS}}$ variables determine which (if
either) of these roots are permitted. In addition to ensuring that the
solutions are real, we must also have that $k^0>\bar{k}^0_3$ and that
$k_3^0>0$. The reality constraint gives
\begin{align}
  \bar{k}^0_3>\frac{1}{2}\Ecm\xi\sqrt{1-y^2}\qquad\Rightarrow\qquad\xi<\frac{2\bar{k}^0_3}{\Ecm}\frac{1}{\sqrt{1-y^2}}\,.
\end{align}
The effect of the remaining two constraints is determined by the sign
of $y$. For $y>0$, only the positive root is a valid solution (since
$k_3^0<0$ for the negative root), and we have a stronger constraint on
$\xi$:
\begin{align}
  y>0:\mbox{positive root for } \xi<\frac{2\bar{k}^0_3}{\Ecm}\,.
\end{align}

For $y<0$, the positive root is valid over the range in $\xi$ set by
the real constraint, and the negative root is valid for
$\xi>2\bar{k}^0_3/\Ecm$:
\begin{align}
  y<0:\,&\mbox{positive root for } \xi<\frac{2\bar{k}^0_3}{\Ecm}\frac{1}{\sqrt{1-y^2}}\\
  &\mbox{negative root for } \frac{2\bar{k}^0_3}{\Ecm}<\xi<\frac{2\bar{k}^0_3}{\Ecm}\frac{1}{\sqrt{1-y^2}}\,.
\end{align}

It remains for us to construct the four momenta of the $\Phi_4$
event. We define
\begin{equation}
  \delta=\frac{k^0}{\bar{k}^0_3}-1\,,
\end{equation}
and the parameter
\begin{equation}
  \beta_{\mathrm{rec}}=\frac{|\vec{\bar{k}}_{\mathrm{rec}}|}{\bar{k}^0_{\mathrm{rec}}}=\frac{\bar{k}^0_3}{\Ecm-\bar{k}^0_3}\,.
\end{equation}
We assign the recoil by defining a boost $\mathcal{B}_t$ along
$\vec{\bar{k}}_{\mathrm{rec}}$ with magnitude $\beta_t$ and a constant
scaling of momenta $\alpha$, so that
\begin{align}
  k_i&=\alpha(\mathcal{B}_t\bar{k}_i)\,,&i&=1,2\,.
\end{align}
Boosting along the recoil direction and then rescaling momenta allows us to keep the recoil three-momentum fixed. We can solve for the parameters $\alpha$ and $\beta_t$ using
\begin{equation}
  k_{\mathrm{rec}}=\alpha(\mathcal{B}_t\bar{k}_{\mathrm{rec}})\,,
\end{equation}
which gives two constraints:
\begin{align}
  k^0_{\mathrm{rec}}&=\alpha\gamma_t\bar{k}^0_{\mathrm{rec}}(1+\beta_t\beta_{\mathrm{rec}})\,,\\
  |\vec{k}_{\mathrm{rec}}|&=\alpha\gamma_t\bar{k}^0_{\mathrm{rec}}(\beta_t+\beta_{\mathrm{rec}})\,.
\end{align}
These can be solved in terms of $\delta$ and $\bar{k}^0_3$ to obtain
\begin{align}
  \label{eq:alphabeta}
  \beta_t&=\frac{\beta_{\mathrm{rec}}^2\delta}{1-\beta_{\mathrm{rec}}^2-\beta_{\mathrm{rec}}\delta}\,,& \alpha &= \sqrt{1-\beta_t^2}\left(1+\frac{\beta_t}{\beta_{\mathrm{rec}}}\right)^{-1}\,.
\end{align}
For the splitting to exist, we must also ensure that
$0<\beta_t<1$. Rewriting $\beta_t$ as
\begin{align}
  \beta_t&=\frac{\bar{k}_3^0(k^0-\bar{k}_3^0)}{\bar{k}^0_3(k^0-\bar{k}^0_3)+\Ecm(k^0_{\mathrm{rec}}-\bar{k}^0_3)}\\
  &= 1-\frac{\Ecm (k^0_{\mathrm{rec}}-\bar{k}^0_3)}{\bar{k}^0_3(k^0-\bar{k}^0_3)+\Ecm(k^0_{\mathrm{rec}}-\bar{k}^0_3)}\,,
\end{align}
we see that for $k^0,k^0_{\mathrm{rec}}>\bar{k}^0_3$ (i.e. for
timelike $k,k_{\mathrm{rec}}$), the condition on $\beta_t$ is
satisfied. Specifically, we require
\begin{equation}
  \bar{k}^0_3<k^0<\Ecm-\bar{k}^0_3\,.
\end{equation}
The upper bound on $k^0$ implies the additional constraint
\begin{equation}
  x<1-\frac{2\bar{k}_3^0}{\Ecm}
\end{equation}
where $x=k^2/\Ecm^2$. This can be translated into a constraint on
$\xi$ and $y$. We may then split $k$ into $k_3$ and $k_4$ using
the FKS variables in the same way as for the FKS splitting.

We can also invert the procedure and construct the projective map from
$\Phi_4$ to $\Phi_3$. Again we must preserve the three-momentum of
the split pair, so that
\begin{align}
  \Phi_{4} :\ & k=(k^0,\vec{k})\,,& k_{\mathrm{rec}}=(\Ecm-k^0,-\vec{k})\,,\\
  \Phi_3 :\ & \bar{k}_3=(|\vec{k}|,\vec{k})\,,&\bar{k}_{\mathrm{rec}}=(\Ecm-|\vec{k}|,-\vec{k})\,.
\end{align}
We need only now define the individual partons in the recoil, which we achieve by using the same boost technique as before. Defining
\begin{equation}
  \bar{k}_i=\mathcal{B}_t^{-1}\left(\frac{1}{\alpha}k_i\right)
\end{equation}
where the inverse boost is now along $-\vec{k}_{\mathrm{rec}}$, as before we can obtain two constraints:
\begin{align}
  \Ecm-|\vec{k}|&=\frac{\gamma_t}{\alpha}(k^0_{\mathrm{rec}}-\beta_t|\vec{k}|)\,,\\
  |\vec{k}|&=\frac{\gamma_t}{\alpha}(|\vec{k}|-\beta_tk^0_{\mathrm{rec}})
\,.
\end{align}
Solving, we naturally recover the same $\alpha$ and $\beta_t$ as in
\eq{alphabeta}. In this case, however, the constraints $0<\beta_t<1$
and $0<\alpha<1$ are automatically satisfied so that the projection
from any $\Phi_{4}$ point onto a $\Phi_3$ point is well-defined.

Finally, we must show that the limits of the splitting Jacobian are indeed equivalent to those in the FKS case. After manipulation of the above expressions, we find that
\begin{align}
  \df\Phi_{4}(q;k1,\dots,k_{4})=J_{\mathrm{FRp}}\df \Phi_3(q;\bar{k}_1,\dots,\bar{k}_3)\df\Phi_{\mathrm{rad}}^{\mathrm{FKS}}(\xi,y,\phi)\,,
\end{align}
where
\begin{equation}
  \df \Phi^{\mathrm{FKS}}_{\mathrm{rad}}=\frac{1}{(4\pi)^3}\df\xi\df y \df\phi
\end{equation}
and
\begin{align}
  \label{eq:frpjac}
  J_{\mathrm{FRp}}=\Ecm^2\xi\left(\frac{2k^0-\Ecm\xi}{2k^0+\Ecm\xi(y-1)}\right)\,.
\end{align}
Substituting \eq{k0sol} into \eq{frpjac}, one can verify that taking
the limit $\alpha\rightarrow 1$ and expanding about $\xi=0$ or $y=1$
one obtains the soft or collinear limits of the FKS Jacobian, see e.g.\
section 5 of \cite{Frixione:2007vw}.
This means that  one can use the same counterterms as those of the FKS
subtraction to obtain a local cancellation of the infrared divergences.

\subsection{Case 2: the emitter as a softer parton}
In the case where the emitter is not the most energetic particle, the
FRp map is no longer appropriate because we no longer need to keep the
thrust-axis aligned with the emitter. In this case, we can use instead
the Catani-Seymour (CS) map \cite{Catani:1996jh}. For example, if we
assume the emitter is $p_2$ and perform the splitting considering
$p_3$ as the spectator parton, the hardest parton $p_1$ is then
unchanged by the splitting and the quantity $(p_2+p_3)^2$ is
preserved.\footnote{When the emitter is instead $p_3$ the r\^oles of the
  emitter and spectator are interchanged but the same quantities are
  preserved.}  This means that the thrust axis remains along
$\vec{p}_1$ and that the value of $\Taufr$ is also unchanged. It is
left for us to show that the singular limit of the Jacobian when using
the CS map with FKS variables is the same as in the FKS case up to an
overall rescaling.

To describe the splitting in this case we adopt the CS notation, where
$\tilde{p}_{ij}$ is the emitter and $\tilde{p}_k$ is the recoil in the
$\Phi_3$ phase space. The daughters of the splitting are labelled
$p_i$ and $p_j$, while $p_k$ is the recoil in the $\Phi_{4}$ phase
space:
\begin{equation}
  \tilde{p}_{ij}+\tilde{p_k}\rightarrow p_i+p_j+p_k=p_{ij}+p_k.
\end{equation}

For the case at hand, we begin by factorising $\Phi_{4}$ into the $4$-parton CS phase space and a radiation part
\begin{equation}
  \df \Phi_{4}=\df\Phi_3^{\mathrm{CS}}(1-y_{ij,k})\Ecm^2\df\Phi_{\mathrm{rad}}(x,\Omega_2)
\end{equation}
where
\begin{equation}
  \label{eq:CSyijk}
  y_{ij,k}=\frac{p_i\cdot p_j}{p_i\cdot p_j + p_i\cdot p_k + p_j\cdot p_k}=\frac{(p_i+p_j)^2}{(p_i+p_j+p_k)^2}
\end{equation}
and $x,\Omega_2$ are a set of variables which parameterise the
splitting $p_{ij}\rightarrow p_i+p_j$. We now wish to express the
$\{x,\cos\theta\}$ in terms of the FKS variables $\{\xi,y\}$ which we
do using the defining relations of the CS and FKS variables:
\begin{align}
  p_{ij}&=\tilde{p}_{ij}+\frac{p_{ij}^2}{2\tilde{p}_{ij}\cdot \tilde{p}_k}\tilde{p}_k,\\
  p_{ij}^2&=x\Ecm^2=2p_i^0p_j^0(1-y),\\
  p_i^0&=\frac{1}{2}\Ecm\xi,\\
  \xi&=\frac{1}{\Ecm}(p_{ij}^0+|\vec{p}_{ij}|\cos\theta).
\end{align}
Solving, we find that
\begin{align}
  \Ecm\xi &= \left(\tilde{p}^0_{ij}+\frac{x\Ecm^2}{2\tilde{p}_{ij}\cdot \tilde{p}_k}\tilde{p}^0_k\right) + \sqrt{\left(\tilde{p}^0_{ij}+\frac{x\Ecm^2}{2\tilde{p}_{ij}\cdot \tilde{p}_k}\tilde{p}^0_k\right)^2 - x\Ecm^2}\cos\theta,\\
  y&=1-x\Ecm^2\left[\Ecm\xi\left(\tilde{p}^0_{ij}+\frac{x\Ecm^2}{2\tilde{p}_{ij}\cdot \tilde{p}_k}\tilde{p}^0_k-\frac{1}{2}\Ecm\xi\right)\right]^{-1}.
\end{align}
The Jacobian of this transformation is given by
\begin{align}
  \frac{\df x\,\df\cos\theta}{\df\xi\df y}\equiv J(x,\cos\theta;\xi,y)=\xi \left(\frac{2p^0_{ij}}{\Ecm}-\xi\right)^2\left(\frac{2\tilde{p}^0_{ij}}{\Ecm}-\xi\right)^{-1}\left(\frac{2|\vec{p}_{ij}|}{\Ecm}\right)^{-1}
\end{align}
so that we have
\begin{align}
  \df \Phi_{4}&=\df \Phi_3^{\mathrm{CS}}(1-y_{ij,k})\frac{2\tilde{p}_{ij}\cdot \tilde{p}_k}{(4\pi)^22\pi}J(x,\cos\theta;\xi,y)\df \xi \df y \df\phi \\
  &\equiv \df \Phi^{\mathrm{CS}}_NJ^{\mathrm{CS}}_{\mathrm{FKS}}(\xi,y)\df \Phi^{\mathrm{FKS}}_{\mathrm{rad}},
\end{align}
and the total Jacobian is
\begin{equation}
  J^{\mathrm{CS}}_{\mathrm{FKS}}(\xi,y)=\left(1-\frac{p_{ij}^2}{2\tilde{p}_{ij}\cdot \tilde{p}_k}\right)\Ecm^2\xi\left(\frac{2p_{ij}^0}{\Ecm}-\xi\right)^2\left(\frac{2\tilde{p}^0_{ij}}{\Ecm}-\xi\right)^{-1}\left(\frac{2|\vec{p}_{ij}|}{\Ecm}\right)^{-1}.
\end{equation}
The soft and collinear limits of this expression are
\begin{align}
  J^{\mathrm{CS}}_{\mathrm{FKS}}(\xi\rightarrow0,y)&=\Ecm^2\xi,\\
  J^{\mathrm{CS}}_{\mathrm{FKS}}(\xi,y\rightarrow1)&=\Ecm^2\xi\left(1-\frac{\xi}{\xi_{\mathrm{max}}}\right),
\end{align}
which are exactly the soft and collinear limits of the Jacobian in the
usual FKS map. Once again, the consequence is that the subtractions
are precisely the same as in the FKS case and we are therefore able to
use the CS mapping consistently with the FKS subtractions.

\subsection{Recasting the mapping for use in the splitting functions}
The mapping which we have constructed in this appendix is used not
only in the fixed-order pieces of \eqs{3masterful}{4masterful} but
also to make the resummed spectrum fully differential in $\Phi_4$ via
the splitting function defined in \eq{fullP}. We must therefore be
able to construct a $\Phi_4$ phase space point using the mapping given
a $\Phi_3$ point and values of three splitting variables.\footnote{This
  is of course also necessary when considering the $2 \to 3 $
  splitting functions. However, since we directly use the FKS map  in
  that case, we do not document here the simpler change of variables
  $\{\Phi_2, \xi, y, \phi\} \to \{\Phi_2, \Tau_2, z, \phi\}$} This can be achieved {\it\`a
  la} FKS, but in order to do so we must express our splitting
variables in \eq{fullP} in terms of the FKS variables.

We consider four momenta $p'_{12},\,p'_3,\,p'_4$ before the splitting
producing a configuration $p_1,\,p_2,\,p_3,\,p_4$ afterwards and
assume the hierarchy $E_1<E_2<E_3<E_4$. Our splitting variables are
defined to be the azimuthal angle $\phi$, the \threej $\Tau_3$ and an
energy ratio $z$:
\begin{align}
  \Tau_3&=2\left(E_1+E_2-\lvert\vec{p}_{12}\rvert\right),\\
  z&=\frac{E_1}{E_1+E_2},
\end{align}
where $\lvert\vec{p}_{12}\rvert=\lvert\vec{p}_1+\vec{p}_2\rvert$.
Rewriting, we have that
\begin{align}
  E_1&=z\left(\frac{1}{2}\Tau_3+\lvert\vec{p}_{12}\rvert\right),\nn\\
  E_2&=(1-z)\left(\frac{1}{2}\Tau_3+\lvert\vec{p}_{12}\rvert\right),\nn\\
  E_{34}&=M_H-\frac{1}{2}\Tau_3-\lvert\vec{p}_{12}\rvert,
\end{align}
while the energy hierarchy which we have assumed limits $z$ to the range
\begin{equation}
  2-\frac{M_H}{\frac{1}{2}\Tau_3+\lvert\vec{p}_{12}\rvert}<z<\frac{1}{2}.
\end{equation}
From the definitions of the FKS variables, we have that
\begin{align}
  \xi&=\frac{2z}{M_H}\left(\frac{1}{2}\Tau_3+\lvert\vec{p}_{12}\rvert\right),\\
  y&=1-\frac{\Tau_3^2+4\Tau_3\lvert\vec{p}_{12}\rvert}{8z(1-z)\left(\frac{1}{2}\Tau_3+\lvert\vec{p}_{12}\rvert\right)^2},
\end{align}
and the Jacobian of the transformation is given by
\begin{equation}
  \label{eq:splittingjac}
J_{\mathrm{split}}=\Biggr\lvert\frac{(1+2\lvert\vec{p}_{12}\rvert)\left[\frac{1}{2}z\Tau_3-(1-z)\lvert\vec{p}_{12}\rvert\right]+\Tau_3\left[1-\Tau_3(1-z)+2z\lvert\vec{p}_{12}\rvert\right]\frac{\partial\lvert\vec{p}_{12}\rvert}{\partial\Tau_3}}{M_Hz(1-z)^2\left(\frac{1}{2}\Tau_3+\lvert\vec{p}_{12}\rvert\right)^2}\Biggr\rvert.
\end{equation}

It remains for us to determine the quantity $\lvert\vec{p}_{12}\rvert$
in terms of the $\Phi_3$ momenta. This depends on whether the FRp or
CS map is being used. In the FRp case, this is rather straightforward
-- the FRp map preserves the value of $\lvert\vec{p}_{12}\rvert$ by
construction and so we have that
$\lvert\vec{p}_{12}\rvert=\lvert\vec{p}\,'_{12}\rvert$. Thus, the last
term in the numerator of \eq{splittingjac} disappears and the
expression simplifies. In the CS case, matters are slightly more
complicated. The map preserves the four-momentum of the most energetic
particle $p_4'$ while $p_3'$ is the spectator; from the definition of
the CS variables, we have that
\begin{align}
  p_4&=p_4',\nn\\
  p_3&=(1-y_{12,3})p_3'.
\end{align}
Using the definition of $y_{ij,k}$ in \eq{CSyijk}, we find that
\begin{equation}
  \lvert\vec{p}_{12}\rvert=\frac{\frac{1}{2}\Tau_3-M_H+E'_4+\left(1-\frac{\Tau_3^2}{4(p'_{12}+p'_3)^2}\right)E'_3}{\frac{E'_3\Tau_3}{(p'_{12}+p'_3)^2}-1},
\end{equation}
where the right hand side is expressed solely in terms of primed
quantities and splitting variables. We may then substitute into
\eq{splittingjac} which provides us with all the information we
require.

\section{NNLO decay rates}
\label{app:NNLOdecrates}
For convenience, we list here the NNLO decay rates of the Higgs boson
to $b$-quarks and to gluons, taken from
\cite{Chetyrkin:1996sr,Schreck:2007um}. In the following, $L_H\equiv
\log(\mu^2/M_H^2)$ and $L_t\equiv\log(\mu^2/m_t^2)$,  we have set
$N_c=3$ and  $n_\ell$ is the number of light active flavours.

\begin{align}
  \Gamma^b_{\mathrm{NNLO}}(\mu)=\Gamma_\mathrm{B}^b(\mu)\Bigg\{1&+\left(\frac{\as(\mu)}{4\pi}\right)\left[\frac{68}{3}+8L_H\right]\nn\\
  &+\left(\frac{\as(\mu)}{4\pi}\right)^2\bigg[\frac{10801}{9}-\frac{76\pi^2}{3}-312\zeta(3)+\frac{1696}{3}L_H+76L_H^2\nn\\
  &+n_\ell\left(-\frac{130}{3}+\frac{8\pi^2}{9}+\frac{32}{3}\zeta(3)-\frac{176}{9}L_H-\frac{8}{3}L_H^2\right)\bigg]\Bigg\}
\end{align}

\begin{align}
  \Gamma^g_{\mathrm{NNLO}}(\mu)=\Gamma_\mathrm{B}^g(\mu)\Bigg\{1&+\left(\frac{\as(\mu)}{4\pi}\right)\bigg[95+22L_H+n_\ell\left(-\frac{14}{3}-\frac{4}{3}L_H\right)\bigg]\nn\\
  &+\left(\frac{\as(\mu)}{4\pi}\right)^2\bigg[\frac{149533}{18}-121\pi^2-990\zeta(3)+3301L_H+363L_H^2+38L_t\nn\\
    &+n_\ell\left(-\frac{8314}{9}+\frac{44\pi^2}{3}+20\zeta(3)-380L_H-44L_H^2+\frac{32}{3}L_t\right)\nn\\
    &+n_\ell^2\left(\frac{508}{27}-\frac{4\pi^2}{9}+\frac{28}{3}L_H+\frac{4}{3}L_H^2\right)\bigg]\Bigg\}
  \end{align}
\bibliographystyle{JHEP}
\bibliography{geneva}

\end{document}